\documentclass[12pt,letterpaper,preprint]{aastex}
\usepackage{amsmath}
\usepackage{amsfonts}
\usepackage{amssymb}
\usepackage{float}
\usepackage{graphicx}
\usepackage{natbib}
\begin{document}

\title{GENERAL RELATIVISTIC HYDRODYNAMIC SIMULATION OF ACCRETION FLOW FROM A STELLAR TIDAL DISRUPTION}

\author{Hotaka Shiokawa\altaffilmark{1}, Julian H. Krolik\altaffilmark{1}, Roseanne M.
Cheng\altaffilmark{1,2},  Tsvi Piran\altaffilmark{3}, and Scott C. Noble\altaffilmark{4,5}}

\altaffiltext{1}{Physics and Astronomy Department, Johns Hopkins University, Baltimore, MD 21218,
USA}

\altaffiltext{2}{Center for Relativistic Astrophysics, School of Physics, Georgia Institute of
Technology, Atlanta, GA 30332, USA}

\altaffiltext{3}{Racah Institute of Physics, The Hebrew University of Jerusalem, Jerusalem 91904,
Israel}

\altaffiltext{4}{Center for Computational Relativity and Gravitation, School of Mathematical
Sciences, Rochester Institute of Technology, Rochester, NY 14623}

\altaffiltext{5}{Department of Physics and Engineering Physics, University of Tulsa, Tulsa OK 74104}

\begin{abstract}
We study how the matter dispersed when a supermassive black hole tidally
disrupts a star joins an
accretion flow.    Combining a relativistic hydrodynamic simulation of
the stellar disruption with a relativistic
hydrodynamics simulation of the tidal debris motion, we track such a
system until $\simeq 80\%$ of the stellar mass
bound to the black hole has settled into an accretion flow.   Shocks
near the stellar pericenter and
also near the apocenter of the most tightly-bound debris dissipate
orbital energy, but only enough
to make the characteristic radius comparable to the semi-major axis of
the most-bound material,
not the tidal radius as previously thought.  The outer shocks are caused
by post-Newtonian
effects, both on the stellar orbit during its disruption and on the
tidal forces. Accumulation of mass into
the accretion flow is non-monotonic and slow, requiring $\simeq 3-10\times$ the
orbital period of the most tightly-bound
tidal streams, while the inflow time for most of the mass may be
comparable to or longer than the mass
accumulation time.  Deflection by shocks does, however, remove enough
angular momentum and
energy from some mass for it to move inward even before most of the mass
is accumulated into
the accretion flow.   Although the accretion rate rises sharply and then
decays roughly as a power-law,
its maximum is $\simeq 0.1\times$ the previous expectation, and the duration of the
peak is $\simeq 5\times$ longer than previously
predicted.  The geometric mean of the black hole mass and stellar mass
inferred from a measured event
timescale is therefore $\simeq 0.2 \times$ the value given by classical theory.
\end{abstract}

\section{Introduction}\label{sec:intro}

When a star comes close enough to a black hole, strong tidal forces can tear off a significant amount
of mass from the star, and possibly tear apart the star completely.  These ``Tidal Disruption Events" (TDEs) are
thought to be common in the centers of galaxies \citep{Komossa.Bade.1999, Wang.Merritt.2004} .
Although quite uncertain, the estimated rate for main sequence stars (MS stars) to be disrupted by supermassive
black holes (SMBHs) is $\sim 10^{-5}$~yr$^{-1}$~galaxy$^{-1}$ \citep{Wang.Merritt.2004,Brockamp.et.al.2011,vanVelzen2014}.
TDEs are also expected to happen when a white dwarf (WD) passes too close to an
intermediate mass black hole (IMBH), possibly in a globular cluster \citep[e.g.][]{KP2011,Haas.et.al.2012,MacLeod2014}
or possibly in a galactic center, but the timescale is shorter than for a main sequence star disruption
by the square root of their mass density ratio.

The closest distance a star in a parabolic trajectory can approach a black hole without being disrupted is
called the tidal radius $R_t \simeq R_{\ast} (M_{BH}/M_{\ast})^{1/3}$, where $R_{\ast}$ is the radius of the
star, $M_{\ast}$ is the mass of the star, and $M_{BH}$ is the mass of the black hole.   More precisely, the tidal
radius can be written \citep{Phinney.1989} as $R_t \simeq 50 (k/f)^{1/6} (M_*/M_{\odot})^{2/3-\xi} M_{BH,6}^{-2/3}R_g$, where
$k$ is the star's apsidal motion constant (a function of its internal density profile), $f$ is its binding energy
in units of $GM_*^2/R_*$, and $R_g \equiv GM_{BH}/c^2$.   The ratio $k/f$ ranges between $\simeq 0.02$ for
fully radiative stars and $\simeq 0.3$ for fully convective stars \citep{Phinney.1989}.   The index $\xi$ parameterizes
the radius-mass relation: with $R_* \propto M_*^{1-\xi}$, $\xi \simeq 0.2$ for $M_* \lesssim M_{\odot}$, but rises to
$\simeq 0.4$ for higher-mass stars \citep{Kippenhahn1994}.
For a complete disruption to happen, the actual pericenter of the star's orbit $R_p$ must be $\leq R_t$, and
$R_t$ must be outside the black hole event horizon.    This last condition
poses an approximate upper limit on the black hole mass: for a MS-SMBH encounter,
$M_{BH} \lesssim 10^8(k/f)^{1/4}(M_*/M_{\odot})^{1 - 3\xi/2}M_{\odot}$ and
for a WD-IMBH encounter, $M_{BH} \lesssim 10^5M_{\odot}(M_*/M_\odot)^{-1}$.   In both cases, the actual order-unity
coefficient of the maximum mass requires a more careful relativistic calculation and generically increases by a factor of
several for black holes with larger spin parameters \citep{Kesden2012}. 

Consideration of these events by a number of authors \citep{Hills75,FR76,Lacy82,Luminet85} led to a concise
description of the fate of stellar debris in a full disruption as determined by its energy distribution \citep{Rees.1988}.
The key---and plausible---supposition is that the energy per unit mass of the different fluid elements torn off the star has a roughly flat
distribution from $\sim - E_{m*}$ to $\sim +E_{m*}$, where $E_{m*}$ is the binding energy per unit mass, with respect
to the black hole, of the matter on the side of the star nearest the black hole.    The energy distribution is symmetric
about zero when, as is generally the case, the star approaches on a parabolic orbit.   One immediate consequence
of this energy distribution is that half the star's mass escapes the potential well of the black
hole, while the remaining half travels outward on highly elliptical orbits but then returns after a considerable delay (a
few months for solar-mass main-sequence stars).    This model predicts that at its peak, the mass-return rate can be
very super-Eddington \citep{Loeb.Ulmer.1997}, and the effective
temperature associated with the bulk of the luminosity is generally in the EUV \citep{Rees.1988}.    Once the
mass-return rate passes the peak, this energy distribution predicts a subsequent mass-return rate that diminishes $ \propto t^{-5/3}$
\citep{Phinney.1989}.   For the last twenty-five years, a lightcurve $\propto t^{-5/3}$ has been taken as
the hallmark of a TDE.

In fact, however, there are numerous reasons why this conclusion may have been overly hasty.   One
\citep{Lodato.et.al.2009,Guillochon.Ramirez-Ruiz.2013} is that realistic stellar structures lead to an energy distribution
that isn't exactly flat, so that the $t^{-5/3}$ variation of the mass-return rate becomes valid only at a late stage
or possibly not at all during the event.    Another \citep{Lodato2011} is that even if
the bolometric lightcurve is $\propto t^{-5/3}$, observations are nearly always done in the optical or
near-UV, i.e., on the Rayleigh-Jeans portion of the thermal spectrum, where the flux depends only linearly
on the effective temperature $T$, not $\propto T^4$, as does the bolometric flux.   The observed lightcurve's
decline should therefore be rather shallower than the decline of the bolometric lightcurve.   Still another is that photon-trapping
during the super-Eddington phase of the accretion may create a shallower relation between the mass accretion
rate and the bolometric luminosity \citep{Krolik.Piran.2012}.

In this paper we wish to raise a more fundamental question.  Identification of the mass-return rate with the
mass accretion rate onto the black hole, where the majority of the power is generated, demands that the time
between the return of a tidal stream and its accretion onto the black hole is short compared to the return time of the stream.
This is not a trivial requirement.  Although all the tidal streams initially have specific angular momentum close to the initial
specific angular momentum of the star, they have orbital energies much larger than that associated with a circular orbit
having that angular momentum.  That is why their orbits are highly elliptical: their semi-major axes are all $\gg R_p$ (by ratios
$\gtrsim (M_{BH}/M_*)^{1/3}$).  To join a disk at a radius $r \sim R_p$ requires that the streams' energy be reduced by this
same large factor.   \cite{Rees.1988} argued that  relativistic apsidal precession would cause
stream intersections near $R_p$ resulting in strong shocks.   As a result of the dissipation of a large part of the
streams' orbital energy, the tidal streams would promptly join an accretion disk whose outer radius is $\simeq 2R_p$;
the inflow time from there to the black hole is less than the characteristic return time.

However, if the elliptical orbits of the tidal streams are all mutually aligned, coinciding only at their pericenters, the only
place where stream orbits intersect is at the pericenter, but the shocks taking place there dissipate a fraction of the orbital
energy that is only $\sim \theta^2$, where $\theta$ is the angular thickness of the stream orbits about the
midplane \citep{Kochanek1994,Guillochon.et.al.2014}.
General relativistic precession of the elliptical orbits' apsidal orientation might help \citep{Cannizzo90,Hayasaki13},
but when the star's pericenter is several tens of $R_g$ or more, the precession angle is not
great enough to engender shocks near $R_p$ strong enough to dissipate a large fraction of the orbital energy;
in addition, as we will show here, although apsidal precession does create shocks, they are near the apocenter
of the orbit, and those shocks produce enough deflection so as to eliminate the contribution of apsidal precession
to shocks near the pericenter.    For similar reasons, Lense-Thirring precession, although very likely present, is
unlikely to create sufficiently strong shocks near the pericenter
\citep{Kochanek1994,Haas.et.al.2012,StoneLoeb2012}.   {\it In other words, we have no mechanism to realize
a key step in the speedy delivery of returning matter to the black hole, the quick dissipation of orbital
energy.}    In fact, the situation is even worse than this because  merely
dissipating orbital energy into heat doesn't solve the problem: it means only that the gas is supported by pressure
rather than rotation while its characteristic distance from the central mass is, to order of magnitude, unchanged.
The heat must be radiated in order for mass to move inward, and given the large optical depth of a star spread
over many tens of black hole gravitational radii, that may not be quick.  It is the goal of this paper to study in detail the
hydrodynamics of the returning tidal streams in order to identify the locations and mechanisms of orbital energy
dissipation, and therefore how those streams may ultimately join an accretion flow onto the central black hole.

Some hydrodynamical simulation studies of these processes have already been carried out
\citep{Rosswog.et.al.2009,Haas.et.al.2012,Guillochon.et.al.2014}, but all previous work has been limited to
durations a small fraction of that required to track the return of the majority of the star's bound mass.
With our technique, it is possible to integrate long enough ($\simeq 12\times$ the orbital period of the most
tightly-bound tidal stream) to follow the return of nearly all the star's bound mass.   By so doing, we can determine
the structure of the system formed upon the streams' return and examine the degree to which it resembles
a classical accretion disk.   Most crucially, we will determine the level of dissipation due to hydrodynamical processes
suffered by the half of the star's mass remaining bound to the black hole.    The process by which the tidal streams
form a disk is often called ``circularization" in the literature, in the expectation that what results is, in fact, a system
in which the fluid follows nearly circular orbits around the black hole.   Another of the goals of this paper is to define
this process in quantitative terms so that progress toward complete ``circularization" can be measured as a
function of time.

To accomplish our goals, we must solve two distinct problems: the internal dynamics of the star
as it falls apart under the tidal stresses of the black hole, and the hydrodynamics of the tidal streams
as they orbit the black hole and interact with one another.     We therefore divide our computational
method into two corresponding sections.   In the first, we follow the star as its self-gravity is overcome
by general relativistic tidal stresses, and internal pressure gradients push its matter outward even as
some portions are compressed.  This calculation is conducted in the frame of motion that the center-of-mass
of the star would follow if it were unaffected by tides; it follows what happens to the star as the star
falls toward the black hole, swings around, and is tidally destroyed.   In the second, we employ a general
relativistic hydrodynamics code to compute the hydrodynamics of the resulting tidal streams as they orbit around
the black hole; this calculation is conducted in the black hole frame.   The initial conditions
for the latter are provided from the results of the former by taking the distribution of energy and angular momentum
ejected from the star across a range of times $t^\prime$ and computing where the associated orbits will reach at a single
later time $t$.   The details of these methods are presented in Section~\ref{sec:SimDetails}.

For technical reasons explained below, the actual case we treat is a white dwarf disrupted by a relatively small ($500M_{\odot}$)
black hole.   Fortunately, many of our results can be scaled to the much more common case of main sequence
stars disrupted by the larger black holes generally found in galactic nuclei.

\section{Simulations}\label{sec:SimDetails}

\subsection{Simulation parameters}

In ``typical" tidal disruptions, one might expect the star to be a main sequence star of $\sim 1 M_{\odot}$
while the black hole has a mass $\sim 10^6 M_{\odot}$ or more.
Unfortunately, when the ratio $M_{\rm BH}/M_*$ is this large, the dynamic range of lengthscales and
timescales makes simulation rather expensive.   The ratio between the semimajor axis of the most
tightly-bound tidal streams and the tidal radius is $\sim (M_{\rm BH}/M_*)^{1/3}$; the ratio between
the gravitational dynamical timescales at the two radii is $\sim (M_{\rm BH}/M_*)^{1/2}$.    For this
reason, we (and a number of previous workers in this field) have chosen to study an encounter between
a white dwarf and a smaller mass black hole.    In particular, we have chosen a white dwarf with mass
$M_\ast = 0.64 M_{\odot}$ and a black hole of mass $M_{\rm BH} = 500M_{\odot}$.   Although arguably
rarer than main sequence star disruptions, white dwarf disruptions are possible and may have been
observed \citep{KP2011}.


We choose the pericenter of the star's orbit $R_p$ to be identical to the tidal radius $R_t$, the distance from the
black hole at which a star can be fully disrupted.   With this definition, $R_p = R_t \sim (M_{\rm BH}/M_*)^{1/3} R_*$,
where $R_*$ is the radius of the star.   For a white dwarf of the
mass we have chosen, $R_\ast = 8.62 \times 10^8$~cm when its internal structure is assumed to be an $n=3/2$
polytrope.   For the black hole mass we have selected, we then find
$R_t = 7.94 \times 10^9$~cm.   In gravitational units (which we will predominantly use), the
fundamental lengthscale is $R_g = GM_{\rm BH}/c^2 = 7.38 \times 10^7 (M_{\rm BH}/500 M_{\odot})$~cm,
making $R_t = 108 R_g$ and $R_* = 11.7 R_g$.    If $G=c=1$, $R_g=M$, an abbreviation we will often use.

Depending on the context, time is best described in any of several units.   In the code, the time unit is the relativistic
one, $R_g/c \equiv M$ when $G=c=1$.    For discussion of tidal stream orbits, the characteristic time is the orbital
period in the black hole's gravity for a test-particle with an orbital energy equal to (minus) the contrast in gravitational
potential energy (with respect to the black hole) from the star's surface to its center when it passes pericenter,
$\Delta \epsilon_N= R_* R_g /R_p^2 = 1.01 \times 10^{-3}$.   This characteristic timescale for our case is
$t_0 = 6.92 \times 10^4M = 170$~s and corresponds to the orbital timescale of the most tightly-bound matter,
whose semi-major axis is $\simeq 500M$.
Time measured in these units will be denoted by $\tau \equiv t/t_0$.
We set the zero-point of time to be the moment at which the star passes through pericenter.

Finally, we choose the spin of the black hole to be 0 for simplicity.    With $R_p \sim 100M$, the
effects of spin would be minor in any case.

\subsection{Stellar disruption and its consequences}

The first step in our calculation is to compute the trajectory of
the star as it passes by the black hole on a marginally bound orbit with pericenter $R_p$.
We do so including all relativistic effects.
The dynamics of tidal disruption are computed in a local Fermi normal coordinate (FNC) frame moving
with this trajectory \citep{Cheng2013,Cheng.Bogdanovic.2014}. 
The tidal field is described by a hexadecapole (including terms up to $l=4$) multipole expansion of the
tidal stress (i.e., curvature tensor) in the FNC system.   It is then expanded to 4th-order in $[R_g/R(t)]^{1/2}$,
where $R(t)$ is the instantaneous distance of the star from the black hole, in order to construct
a 2PN (second-order post-Newtonian) approximation to the relativistic tidal stress.    
Within a cubical box  $8R_*$ on a side, we use a version \citep{Cheng2013} of the intrinsically conservative
grid-based code VH1 \citep{Blondin2012} to compute the Newtonian self-gravity of the star
and the Newtonian hydrodynamics induced by the combination of stellar self-gravity and
the 2PN approximation to the relativistic tidal stresses.    This simulation extends from a time several dynamical timescales
before pericenter passage until several dynamical timescales after pericenter passage.

\begin{figure}
\begin{center}
\includegraphics[width=140mm]{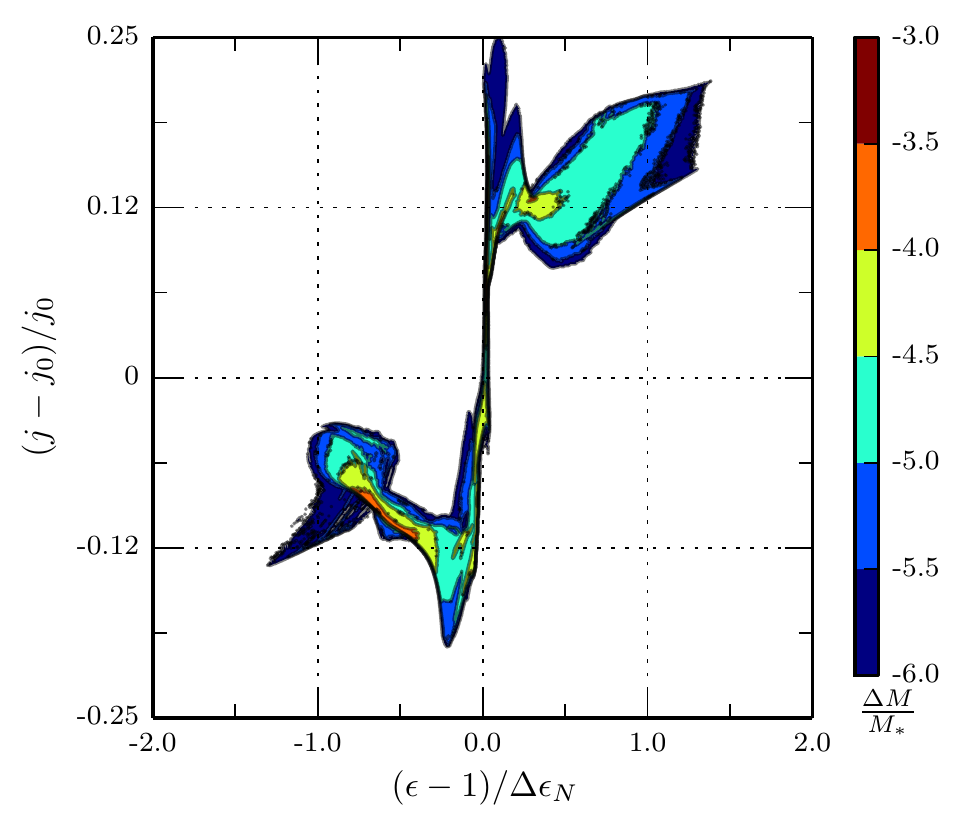}
\end{center}
\caption
{Distribution of mass with respect to specific angular momentum (vertical axis)
and specific orbital energy (horizontal axis).   $\epsilon - 1$ is the energy per
unit mass after subtracting the rest-mass; $\epsilon - 1<0$ is bound, $\epsilon - 1 > 0$
is unbound.    The color scale for mass is logarithmic.
}
\label{fig:LEdistfunct}
\end{figure}

Throughout this time, we calculate the rate at which mass leaves the box with varying
amounts of momentum and internal energy.    Figure~\ref{fig:LEdistfunct} shows the
distribution function for all ejected mass in the orbital energy--angular momentum plane.
The great majority of the matter bound to the black hole has specific angular momentum $j$ in the
range 0.86--$0.93 \times$ the specific angular momentum of the star, $j_0$; the unbound
matter distribution is similarly centered on $\simeq 1.1 \times j_0$.   A small amount of mass
has specific orbital energy  as small  as $\simeq -1.3 \Delta \epsilon_N$, but there are significant
amounts only for $\epsilon - 1 \gtrsim - 0.9 \Delta\epsilon_N $ ($\epsilon$ is the relativistic energy per unit rest-mass;
$\epsilon - 1$ is therefore the Newtonian specific orbital energy).    If this bivariate mass
distribution is integrated over $j$, the mass per orbital energy is, as guessed by
\cite{Rees.1988}, nearly flat across the range $-0.9\Delta\epsilon_N \leq \epsilon - 1 \leq \Delta \epsilon_N$.

The fact that specific energy and angular momentum are so well correlated fits well with the classical
picture that the width of the energy distribution is set by the width in potential energy with respect
to the black hole.   Matter coming from the side of the star nearer the black hole is both deeper in the
black hole's potential and has lower angular momentum.   The contrast in energy per unit mass is, by definition,
$\sim  \Delta \epsilon_N \sim R_* R_g/R_p^2$; the associated fractional contrast in angular momentum per unit mass is
$\sim R_*/R_p$, here $\sim 0.1$, in excellent quantitative agreement with our detailed calculation.

\section{Initialization of the global simulation: mass distribution, orbital shapes, and simulation details}

One of the unique features in this study is that the initial condition for the global simulation
is constructed using a highly-accurate tidal disruption simulation.   Here we describe how
the information is conveyed from one simulation to the other.

Throughout the disruption simulation, mass is ejected through the simulation box edges.
Every $106M$ in time ($\simeq 0.1 \times$ the dynamical time at pericenter), we construct
$6.4 \times 10^5$ pseudo-particles to represent the distribution in momentum
and internal energy of the mass ejected from the star, and injected into the domain of the global
simulation, at that instant.   Each particle is placed in a cell in $(j - j_0)/j_0 \times (\epsilon-1)/\Delta\epsilon_N$
space.    The mass, momentum, kinetic energy, and internal energy of all fluid elements contributing to
each of these cells are summed; from these, the pseudo-particle is given an entropy in the sense of $p/\rho^\gamma$.
Under the assumption that the propagation of each pseudo-particle is entirely ballistic, we project its
location along the geodesic from its injection point to its predicted location
at time $\tau = 0.62$ after the star passes pericenter.

At the end of the local simulation, we determine how many particles carrying how much mass,
momentum, orbital energy, and entropy are located in each global simulation cell.
More than $3 \times 10^7$ particles are used in the process so that every simulation cell in the
path of the tidal streams has enough particles to make its average properties well-determined.
The local rest-mass density in a cell is the total mass of particles it contains divided by 
its volume; the local momentum density is computed similarly from the sum of the particle
momenta in the cell.   Making use of the new density, we
convert the mean entropy back into internal energy.  This information is
then used as the  initial data for a global, fully general relativistic
hydrodynamics simulation in the black hole frame.

\begin{figure}
\begin{center}
\includegraphics[width=140mm]{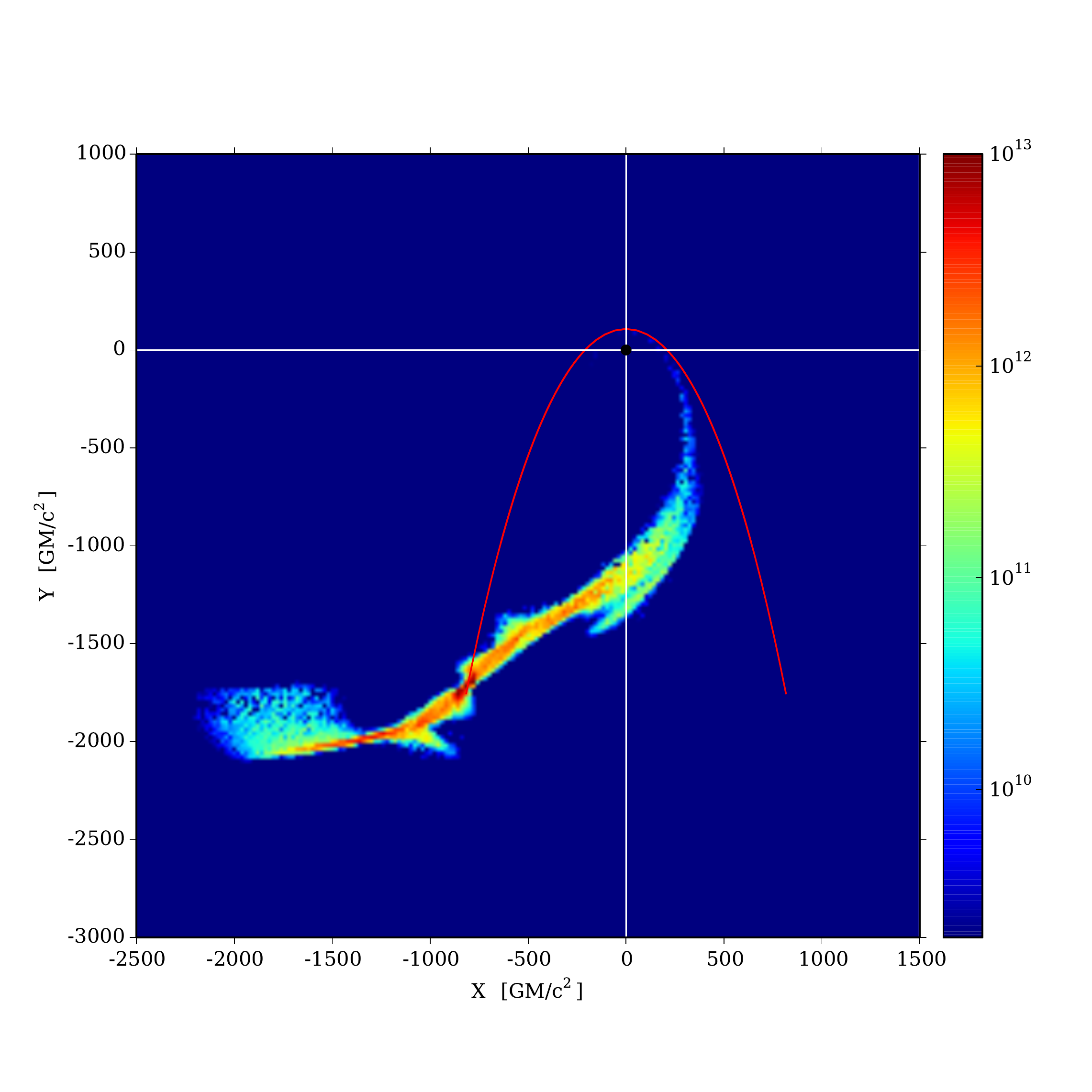}
\end{center}
\caption
{
Initial surface density (in gm~cm$^{-2}$) for the global simulation.
The black hole and the stellar pericenter are at $(X,Y)=(0,0)$ and $(X,Y) \simeq (0,107)$, respectively.
The red curve marks the path of the star's center of mass as it swung counter-clockwise around the
black hole; the gas orbits in the same sense.
}
\label{sigma_init}
\end{figure}

The surface mass density resulting from this initial data is shown in Figure \ref{sigma_init}.
In the figure, bound gas orbits in a counter-clockwise direction.   The coordinates $(X,Y)$ are
Cartesian coordinates whose origin is at the black hole position and are in units of $R_g$.
The core of the original star is at the densest point in the plot, $(X,Y) \simeq (-800M,-1700M)$.
The radius corresponding to that point ($r \simeq 2000M$) roughly
marks the boundary between bound debris (inside) and unbound debris (outside).   
The ratio of the bound to the unbound mass is $\sim 0.46:0.54$.


In order to check the quality of the ballistic transport assumption and our subsequent neglect of
tidal stream self-gravity, we computed three measures of the tidal streams' vertical structure at
early times in the simulation: their actual vertical scaleheight $h$; the hydrostatic scaleheight due
to the black hole's gravity, $h_{\rm BH}$; and the hydrostatic scaleheight due to self-gravity $h_{\rm sg}$.
The latter was computed assuming that the gravity due to the gas's own mass was $4\pi G\Sigma$,
which should be a reasonable estimate when, as is often the case, the horizontal width of the stream
is $> h$.   We find that $h_{\rm BH} \gg h_{\rm sg} \gtrsim h$.    In other words, tidal stream self-gravity
near its initial position is much stronger than the vertical component of the black hole's gravity, but,
somewhat fortuitously, our streams begin as thin or thinner than self-gravity would produce.    This
structure is transient---the gas's pressure expands it vertically---but the vertical dynamical time when
only black hole gravity is present is the same as the orbital dynamical time, so the streams travel a significant portion of
an orbit before they exceed $h_{\rm sg}$ in thickness.   By that time, they encounter the shocks
described later in this paper and become hot enough that vertical self-gravity is no longer significant.

\subsection{Mass Return Rate}\label{sec:mass_return}

\begin{figure}
\begin{center}
\includegraphics[width=140mm]{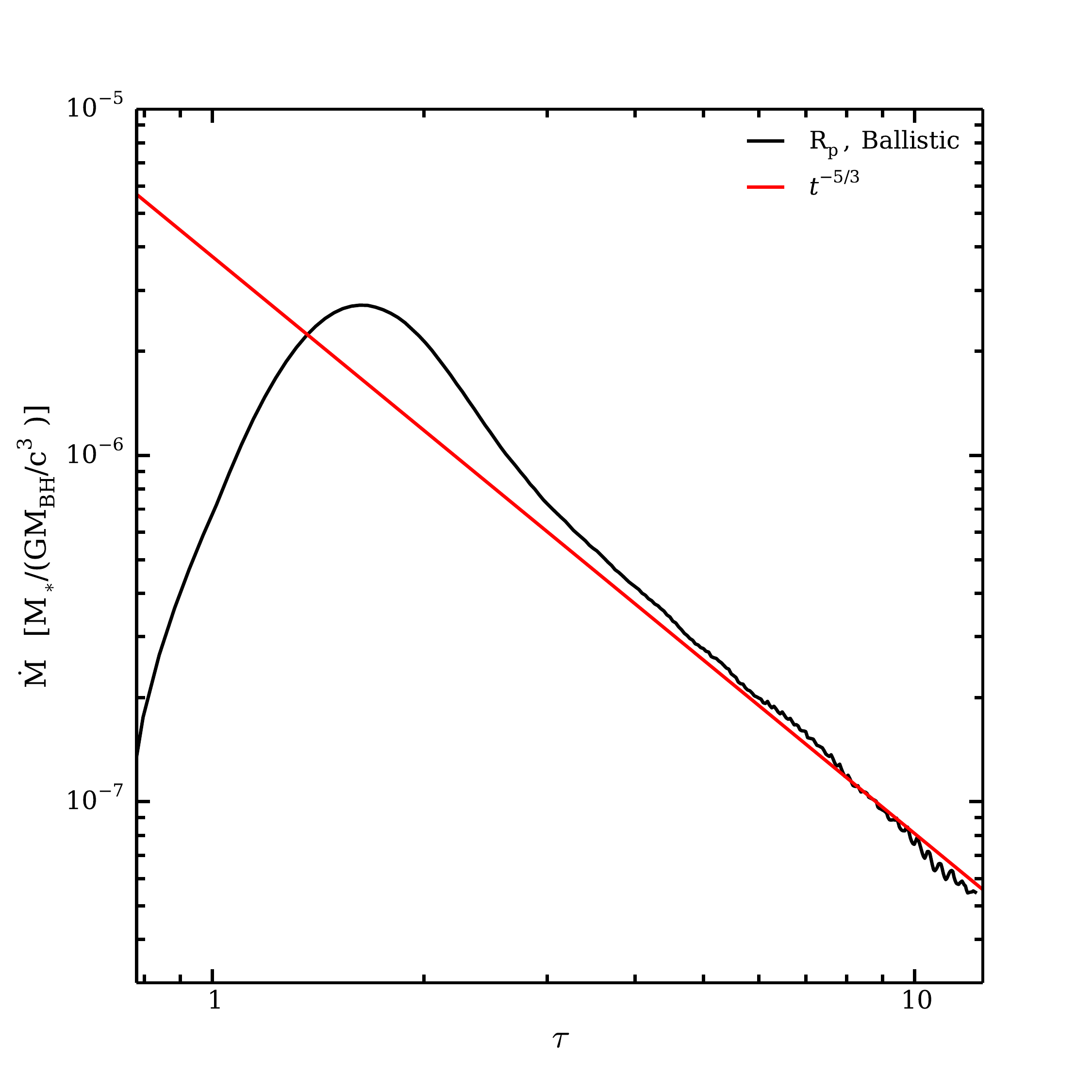}
\end{center}
\caption
{
The black curve is the mass-return rate predicted by the pseudo-particle orbits, i.e., without allowance
for any hydrodynamics.  The red curve is $\propto t^{-5/3}$ and is normalized so that the black curve asymptotes to it.
}
\label{mass_returning_rate}
\end{figure}

As we mentioned in the Introduction, a central prediction of the classical theory of tidal disruption
events is that mass should return in the bound tidal streams at a rate that depends on time $\propto t^{-5/3}$.
We can check the quality of this estimate by following the orbits of our pseudo-particles until all of
them have returned to the pericenter at least once.    In reality, this mass return rate is substantially modified
by hydrodynamics, as our simulation explores, but it is useful as a standard of comparison.
Figure \ref{mass_returning_rate} shows the pseudo-particle prediction from our simulation of the stellar disruption;
at late times it does indeed follow the  $t^{-5/3}$ power law.
However, those ``lates times" don't begin until $\tau \simeq 3$; by this time, more than 60\% of
the bound mass has already returned to the pericenter.    Thus, the time-dependence of even the
mass-return rate can deviate substantially from the usual expectation. 

It is also worth noting that the return rate at the beginning of the global simulation, $\tau=0.62$, is negligibly
small.   This fact supports our choice of this time as the starting point for our hydrodynamic calculation
of the later evolution of the system.

Our result differs in several ways from the predictions made on the basis of analytic estimates
of the specific energy distribution of the tidal streams \citep{Lodato.et.al.2009}.     They found that the
return rate peaks at what we call $t_0$ or $\tau = 1$; we find the peak is at $\tau \simeq 1.5$.     Perhaps more
significantly, we find that the rate of return at that peak lies {\it above} the rate predicted at that time
by an extrapolation of the $t^{-5/3}$ dependence at late times, whereas it lies {\it below} that extrapolation
in the earlier analysis.    On the other hand, we find (after a suitable rescaling to account for the different
parameters) rather better agreement with the return rate computed by \cite{Guillochonweb}, both in terms
of the peak time and the shape of the curve; the only contrasts are that our time of peak return rate is
slightly later than theirs, and the peak return rate we find is a bit higher above the $t^{-5/3}$ curve.

\subsection{Mass Distribution in Apsidal Angle}

\begin{figure}
\begin{center}
\includegraphics[width=140mm]{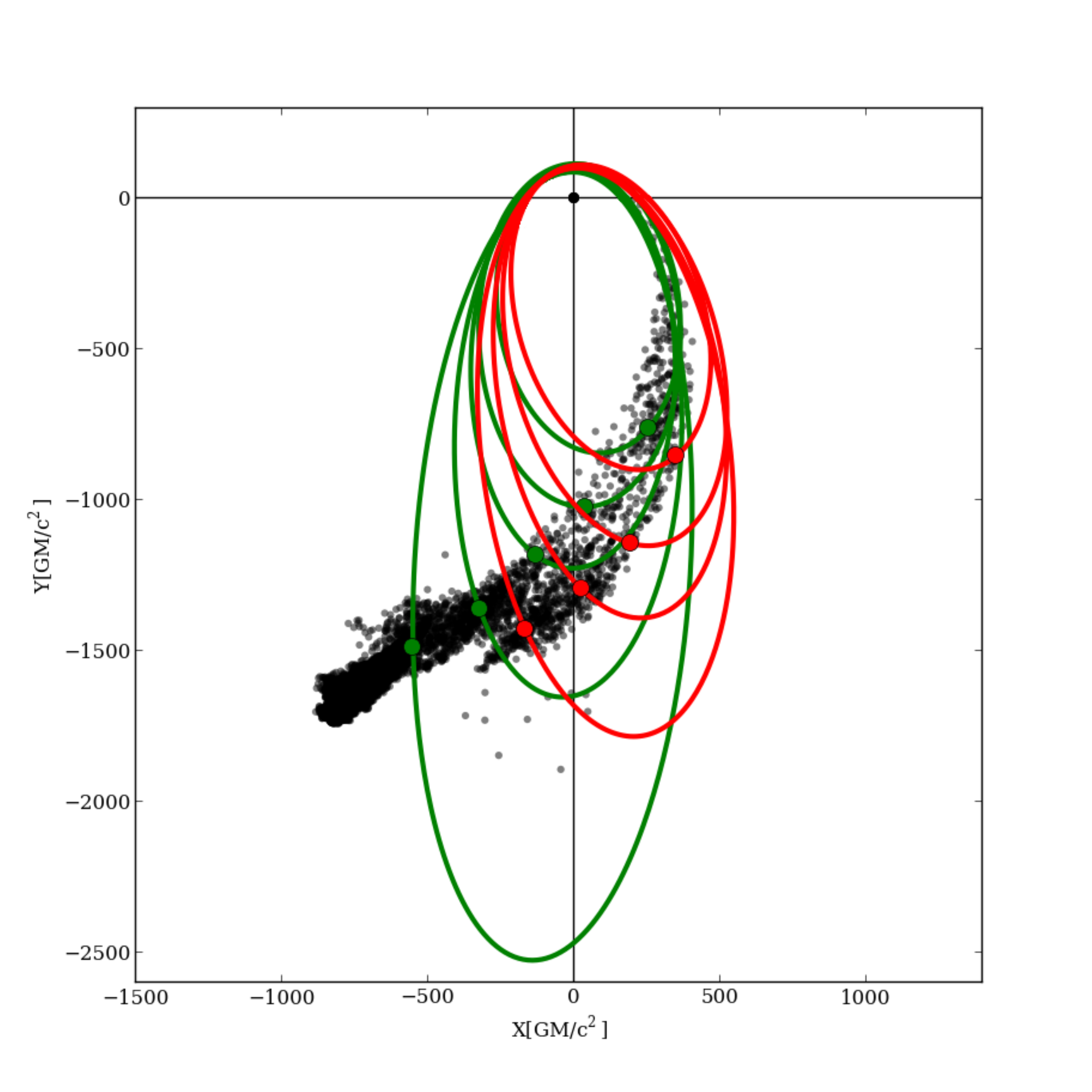}
\end{center}
\caption
{Black points are a selection of pseudo-particles randomly chosen from the bound set to represent
gas density in the global simulation's initial state.   Colored ellipses are single-orbit trajectories
for a small sub-sample of the black points; the starting point for each orbit is a point of matching color.
Red indicates mass that left the star relatively early in the disruption; mass that left the star relatively late
is shown in green.}
\label{fig:traj_init}
\end{figure}

One of the most important properties of the initial condition affecting subsequent global
evolution of the gas is its spread in apsidal angle of the gas orbits.
Figure~\ref{fig:traj_init} shows a single orbit from one pericenter passage to the next for several pseudo-particles.
Unlike the many Newtonian models of post-disruption debris evolution (e.g., \cite{Guillochon.et.al.2014}), our calculation
indicates that the orbital apsidal angles for streams with different semi-major axes are in
general not aligned.    There is a swing of $\simeq 25^\circ$ between streams created early in the encounter
(red in the figure) and streams created later (green in the figure),
comprising the greater part of the mass.    Streams leaving the star early tend to have small
semi-major axes; those leaving later have a broader distribution of semi-major axis.
This misalignment of orbits induces collisions between the gas streams where their orbits intersect;
earlier-arriving gas (smaller semi-major axis) passes the pericenter, heads out
towards its apocenter, and strikes later-arriving gas.   Qualitatively similar rotations in apsidal angle
were seen in the similarly relativistic calculations of \cite{Cheng.Bogdanovic.2014}, which treated
disruptions of main sequence stars by larger black holes.

\begin{figure}
\begin{center}
\includegraphics[width=140mm]{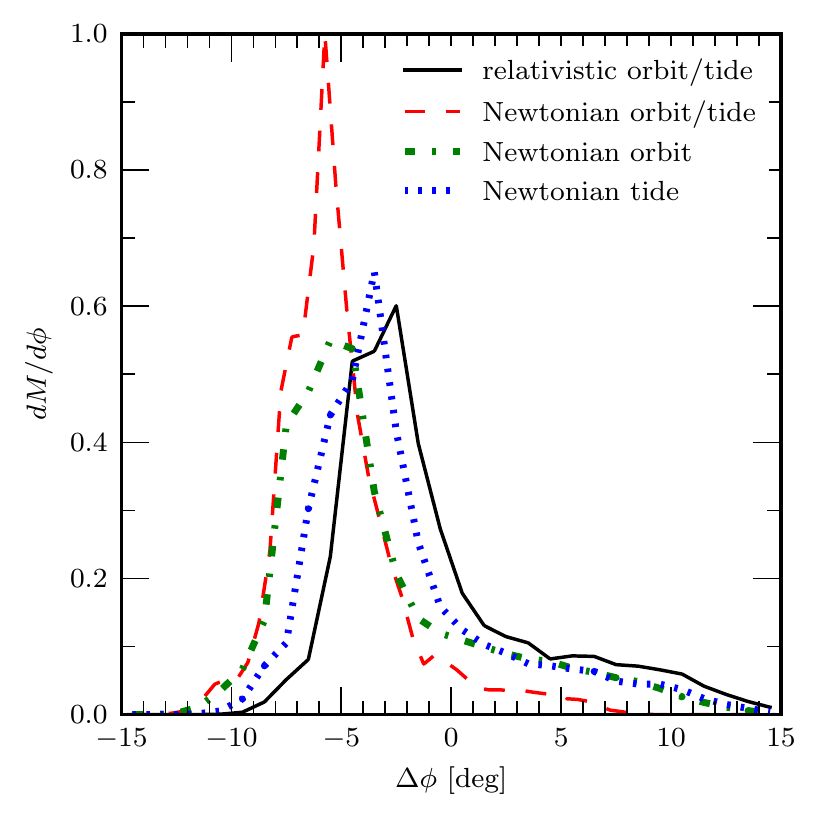}
\end{center}
\caption
{
Distribution of mass per apsidal angle in the initial condition.   The black curve shows the result of
our fully relativistic calculation; the blue curve shows what happens if the tidal stress is described
in terms of Newtonian gravity rather than relativistic but the star's orbit is relativistic; the green curve shows the distribution if the
tidal stress is relativistic but the star follows a Newtonian orbit; the red curve illustrates what happens
in purely Newtonian dynamics.   Negative $\phi$ indicates a tidal stream pericenter before (in the orbital sense)
the star's pericenter; positive $\phi$ after the star's pericenter.
}
\label{fig:dmdphi_phi}
\end{figure}

Figure~\ref{fig:dmdphi_phi} illustrates the distribution of mass with apsidal angle.   We choose
a sign convention in which negative $\phi$ indicates a tidal stream pericenter before (in the orbital sense)
the star's pericenter; positive $\phi$ after the star's pericenter.  In the initial condition, the peak of the
distribution, at $-3^\circ$, is moderately narrow ($\simeq 4^\circ$ FWHM), but the tail to positive $\phi$
extends to such large angles (up to $+15^\circ$) and has such large amplitude that the mass
fraction in the peak is only $\simeq 75\%$, while a full 25\% of the mass is in the tail ($\phi \geq +1^\circ$).    With
$\sim 10^\circ$ contrasts in apsidal direction common,  the angle between many possible orbital pairs at their
near-apocenter intersection is close to $90^\circ$.

The spread in apsidal angles is, in large part, a relativistic effect.   Figure~\ref{fig:dmdphi_phi} also
presents the distributions resulting from treating different portions of the tidal
disruption in Newtonian terms.    A purely Newtonian treatment results in a distribution whose shape
is similar to that given by a relativistic treatment, but shifted $\simeq 3^\circ$ toward more negative apsidal
angles; more importantly, it diminishes the mass fraction in the tail toward larger apsidal angles to a bit less than $4\%$.
Mixed Newtonian/relativistic treatments give intermediate results.   The most significant contributor to
the Newtonian/relativistic contrast comes from the relativistic apsidal precession of
the star's orbit as it swings through pericenter; the change in apsidal direction per pericenter
passage is $3\pi R_g/R_p$ for a highly eccentric orbit, which is $\simeq 6^\circ$ for our
parameters.    Although modified by hydrodynamic effects, each later passage through the pericenter
region results in additional apsidal precession.

The net result of these post-Newtonian relativistic effects is to create a strong correlation between particularly
small semi-major axis, i.e., especially tightly-bound and therefore early-returning matter, and a positive shift in
pericenter direction.


\subsection{Global simulation details}\label{sec:globalgrid}

We use the fully conservative general relativistic magnetohydrodynamics code HARM3D for our global
simulation \citep{Gammie.et.al.2003,Noble.et.al.2009}; there is, however, no magnetic field in the
simulation reported here.    The code uses piecewise parabolic reconstruction, second-order Runge-Kutta time stepping,
and the Lax-Friedrichs flux to solve local Riemann problems.   It is therefore second-order accurate in
smooth flows.

We use modified cylindrical coordinates in Schwarzschild spacetime, where the modified spatial coordinate
variables $(x_1,x_2,x_3)$ are related to ordinary cylindrical coordinate variables $(r,z,\phi)$ by
\begin{align}
r &= \exp{x_1} \\
z &= z_0 \sinh{x_2} \\
\phi &= x_3 \; \text{.}
\end{align}
The simulation volume encompasses $r_{\rm in} = 30M \leq r \leq r_{\rm out} = 5000M$, $0 \leq \phi \leq 2\pi$,
and $-1000M \leq z \leq +1000M$.     Outflow boundary conditions are imposed on all the outer boundaries
of the volume.   We do not treat the region inside $r_{\rm in}$ because we expect that
in real tidal disruptions, MHD effects not treated here will control the system's evolution this close to
the black hole, whereas at larger distances hydrodynamics should be more important.    The outer boundary
is chosen so that $\simeq 80\%$ of the bound mass stays within this radius during the simulation.
The role of the modified coordinates is to place the grid-cells, 192 cells spaced at equal intervals on each of
the modified coordinate axes, where they are most needed within
this volume.   The radial grid cells have constant $\Delta r/r$, but the sinh function in the definition of $x_2$ strongly
concentrates the vertical cells toward the midplane.   The degree of vertical concentration is changed during
the course of the simulation:  $z_0$ increases from 0.59 at the start to 2.94 at $\tau = 2$, and finally to 21.08
at $\tau = 5$.   The midplane cell height therefore changes from $\Delta z= 0.05M$ to $0.2M$ to $1M$. 
Conservation of mass, momentum, and energy is maintained during each grid transition.

The code units of density and internal energy are chosen so that the maximum value in the initial condition for each
is unity.   The unit of all speeds, including sound speeds, is $c$.  In these units, our density and internal energy
``floor values" are $10^{-16}(r/M)^{-3/2}$ and $10^{-20} (r/M)^{-5/2}$, respectively.

With the zero-point of time when the star passes pericenter, the global simulation is terminated at
$\tau = 12.2 \sim 35$min (its actual duration is $\simeq 11.6 $ because it doesn't begin until $\tau = 0.62$).
By the end of the simulation, $81\%$ of the bound mass has returned to the pericenter at least once.

We assume that the gas is adiabatic throughout the simulation.    This is an excellent approximation
for a WD tidal disruption because the cooling time is extremely long: approximately 80~yr, far longer
than the simulation duration.    However, the adiabatic index changes with temperature as radiation
pressure becomes more important relative to gas pressure.   We account for this change by employing the
equation of state in the following way.

Our reasoning begins with the fact that the total pressure
$p=p_{gas}+p_{rad}=(\gamma-1)u_{gas}+u_{rad}/3$ where $p_{gas}$, $p_{rad}$, $u_{gas}$, $u_{rad}$,
and $\gamma$ are gas pressure, radiation pressure, gas internal energy, radiation internal
energy, and gas adiabatic index, respectively.   We can then write
\begin{equation}
p = (\Gamma-1)u
\end{equation}
where
\begin{equation}\label{Gamma}
\Gamma = \gamma - \frac{\gamma-4/3}{1+u_{gas}/u_{rad}} 
\end{equation}
is an ``effective adiabatic index'' and $u=u_{gas}+u_{rad}$ is the total internal energy.
The ratio of gas and radiation internal energy $u_{gas}/u_{rad}$ in equation \eqref{Gamma} is a
function of temperature $T$, i.e. $u_{gas}=nkT/(\gamma-1)$ and $u_{rad}=aT^4$ where $n$, $k$, and $a$ are
the particle number density, the Boltzmann constant, and the radiation constant, respectively.
One can then derive $\Gamma$ for a given pair $(n,u)$ by solving the following equation
for $T$:
\begin{equation}
p_{gas} = nkT = (\gamma-1)u_{gas} = (\gamma-1)(u-u_{rad}) = (\gamma-1)(u-aT^4) \; \text{.}
\end{equation}
If $\gamma$, the gas adiabatic index, is set to 5/3, $\Gamma=5/3$ for $u_{gas}>>u_{rad}$ and $\Gamma=4/3$ for
$u_{gas}<<u_{rad}$.   We find in practice that $\Gamma$ is so slowly-varying that it is unnecessary to account
for its cell-to-cell contrast in the Riemann problem solutions.

\section{Results}

\subsection{Overview}

Perhaps the principal result of our simulation is that tidal streams do {\it not} quickly and easily
join an accretion disk at $r \sim R_p$ immediately upon their return.   Instead, hydrodynamic
effects only gradually ``circularize" their orbits, and the majority of the mass settles at radii
considerably larger than $R_p$.

\begin{figure}
\begin{center}
\includegraphics[width=140mm]{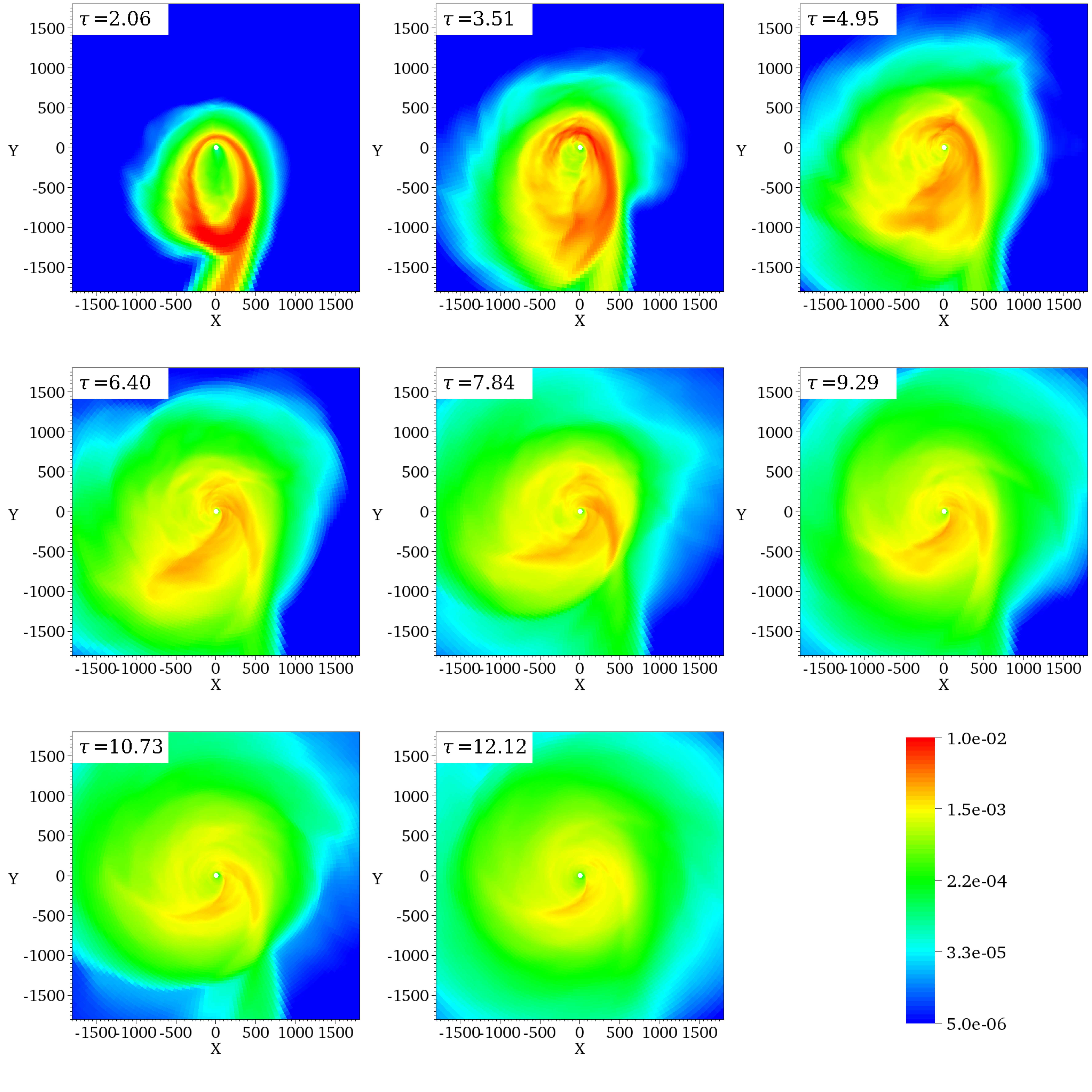}
\end{center}
\caption
{
Time evolution of surface density from $\tau=2.06$ to the end of the simulation, $\tau=12.12$.
The time interval between snapshots is $\simeq 1.5$ in these units.    The spatial axes are in units
of $M$ and are centered on the black hole.   The surface density scale is logarithmic in code units, with a
factor of 2000 from greatest surface density to least.    One code unit of surface density is
$1.11 \times 10^{13}$~gm~cm$^{-2}$.   Note that the highest surface density locations in the
$\tau = 2.06$ panel can be as much as two orders of magnitude greater than the maximum value
of the color scale.   Such large surface densities are not present at later times.
}
\label{fig:sigma_rho_evol}
\end{figure}

\begin{figure}
\begin{center}
\includegraphics[width=140mm]{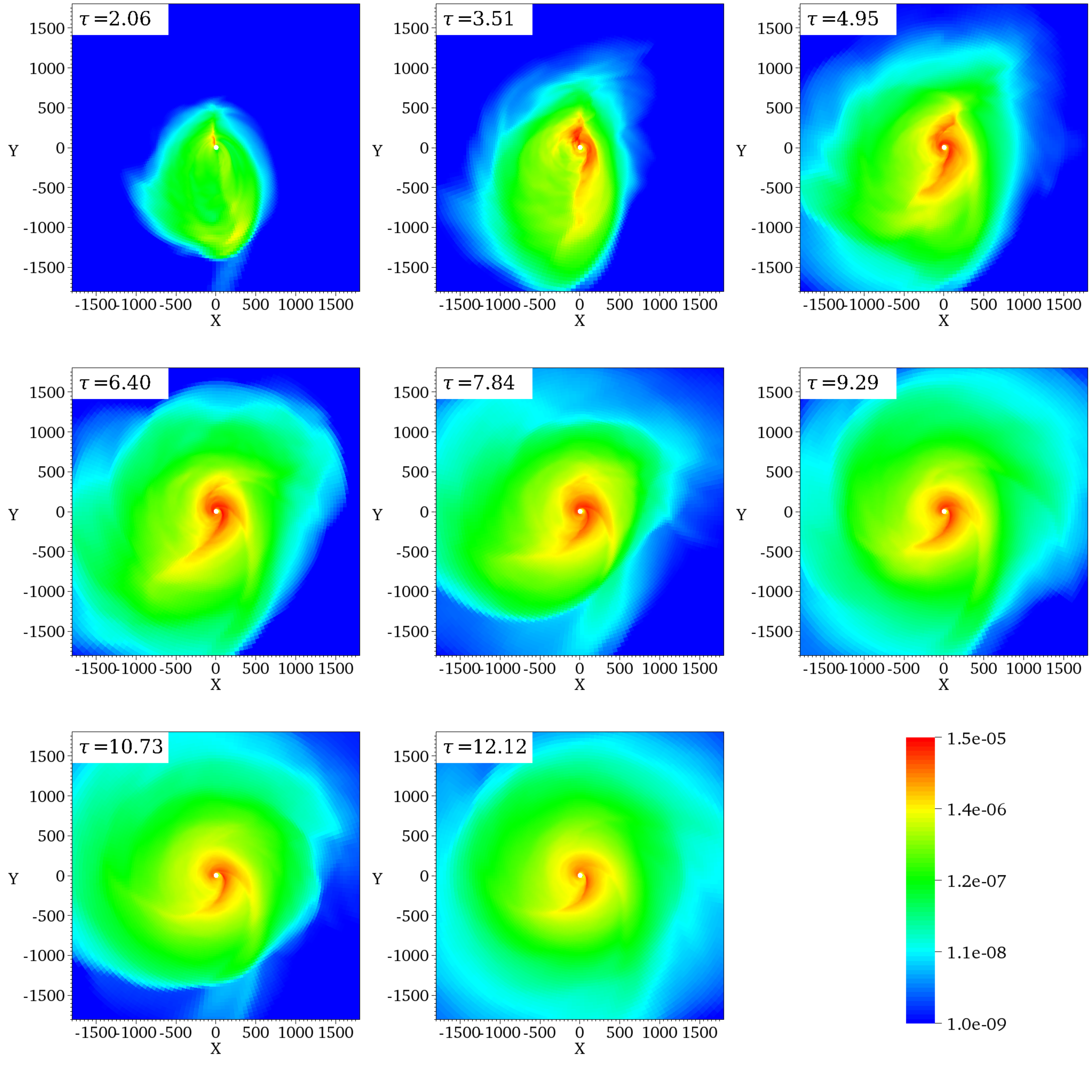}
\end{center}
\caption{
Time evolution of internal energy surface density at the same times as the corresponding panels in
Fig.~\ref{fig:sigma_rho_evol}.    The spatial axes are in units
of $M$ and are centered on the black hole.   The internal energy surface density scale is logarithmic in
code units, with a factor of $1.5 \times 10^4$ from greatest internal energy surface density to least.   One code unit
of energy surface density is $9.96\times 10^{33}$~erg~cm$^{-2}$.
}
\label{sigma_u_evol}
\end{figure}

Figures~\ref{fig:sigma_rho_evol} and \ref{sigma_u_evol}
show the time evolution of surface mass density and surface internal energy density.  
At early times, the surface mass density largely reflects the paths of the tidal streams, but clearly displays a point
of intersection near the streams' apocenter.   As time goes on, the tidal stream features spread and lose contrast.
As they do, the radial profile of the surface mass density reaches a slowly-evolving state in which the azimuthally-averaged
surface mass density is approximately $\propto r$ for $r \lesssim 150M$, is roughly flat out to $r \simeq 600M$, and
then declines steeply at larger radii.  Through much of this period, there are two prominent
ridges of higher density (most noticeable in the panels corresponding to $5 \lesssim \tau \lesssim 11$).   The
origin of these features becomes clearer when the surface mass density and surface internal energy density are viewed
together.

The surface density of internal energy is particularly useful for identifying shock locations, although not all regions of
elevated internal energy density contain shocks.   For example, early in the evolution ($\tau \simeq 0.8$), a shock appears
close to the stellar pericenter $(X,Y) \simeq (0,100M)$; this shock can be seen clearly as the bright spot just above the
black hole in the $\tau = 2.06$ panel of Fig.~\ref{sigma_u_evol}.   The result of vertical gravitational focusing \citep{Kochanek1994},
this shock is often called the ``nozzle shock"  \citep{Evans.Kochanek.1989,Guillochon.et.al.2014}, we call it ``shock~1".
As we will show later, hydrodynamic effects alter the structure and even location of shock~1 over time as some
material makes multiple passes through this region (see Fig.~\ref{schematic} for a pedagogical sketch of
this and other shocks' evolution).

Shortly after the formation of shock~1, an outer  system of shocks forms at large radius ($r \lesssim 1000M$) where
gas flowing outward after passing through the pericenter region collides with newly returning gas.  Although visible in the
$\tau = 3.51$ panel, these shocks are more easily seen beginning at $\tau = 4.95$: they are the two legs of the
feature resembling a ``$\lambda$".    As can be seen in Figure~\ref{fig:traj_init}, in the limit of very narrow stream
width, the only shock created in a set of perfectly-aligned orbits would be the nozzle shock; outer shocks appear
only as the result of finite stream width, or, more powerfully, as the result of rotation in apsidal angle.
Much as in the case of supernova remnants,
the encounter between gas on its way out and newly returning gas creates a ``forward" and a ``reverse" shock,
which we call, respectively, shock~3 and shock~2 (a shock in roughly the same location as shock~2 was
noted by \citet{Rosswog.et.al.2009}).    Material falling in from large radius encounters the forward shock,
shock~3; shock~2 acts upon material that has already taken at least one turn around the black hole.    Initially, shock~2
has only one segment, located very close to shock~3 (see the $\tau = 2.06$ and $\tau = 3.51$ panels of
Fig.~\ref{sigma_u_evol}; the faint red feature at large negative $Y$ and small positive $X$ marks these two
shocks).   However,  as gas accumulates behind it, shock~2 splits (see the panels for $\tau=4.95$--10.73).
One branch stays at roughly the same location and remains very close to shock~3, but a new branch moves
upstream, swinging $\sim 1$~radian clockwise in
azimuthal angle from the stationary branch (see any of the later panels in Fig.~\ref{sigma_u_evol}).    The early
appearance and, to some degree, the very existence, of shocks~2 and 3 is a consequence of the
rotation in apsidal angle of the more tightly-bound tidal streams.   

All of the shocks broaden the fluid's specific angular momentum distribution by deflecting streams away
from their ballistic orbits;  in these deflections, some streams lose angular momentum while others gain it.
Those losing angular momentum move inward on more eccentric orbits, while those gaining angular momentum
traverse less eccentric orbits with larger pericenters.   Although, as we will discuss in greater detail in
Sec.~\ref{sec:circularization}, this process creates more circular orbits for most of the gas mass,
it does not involve significant orbital energy dissipation or lead directly to matter joining an accretion
disk at radii near $R_p$.    In fact,  after the bulk of the mass has returned ($\tau \sim 4$), the surface density
peak gradually shifts outward from the vicinity of the stellar pericenter to $\approx 3 R_p$; this shift is
due to the passage through the inner boundary of matter that has lost angular momentum, so that
the mean angular momentum in the matter left behind increases.

\begin{figure}
\begin{center}
\includegraphics[width=140mm]{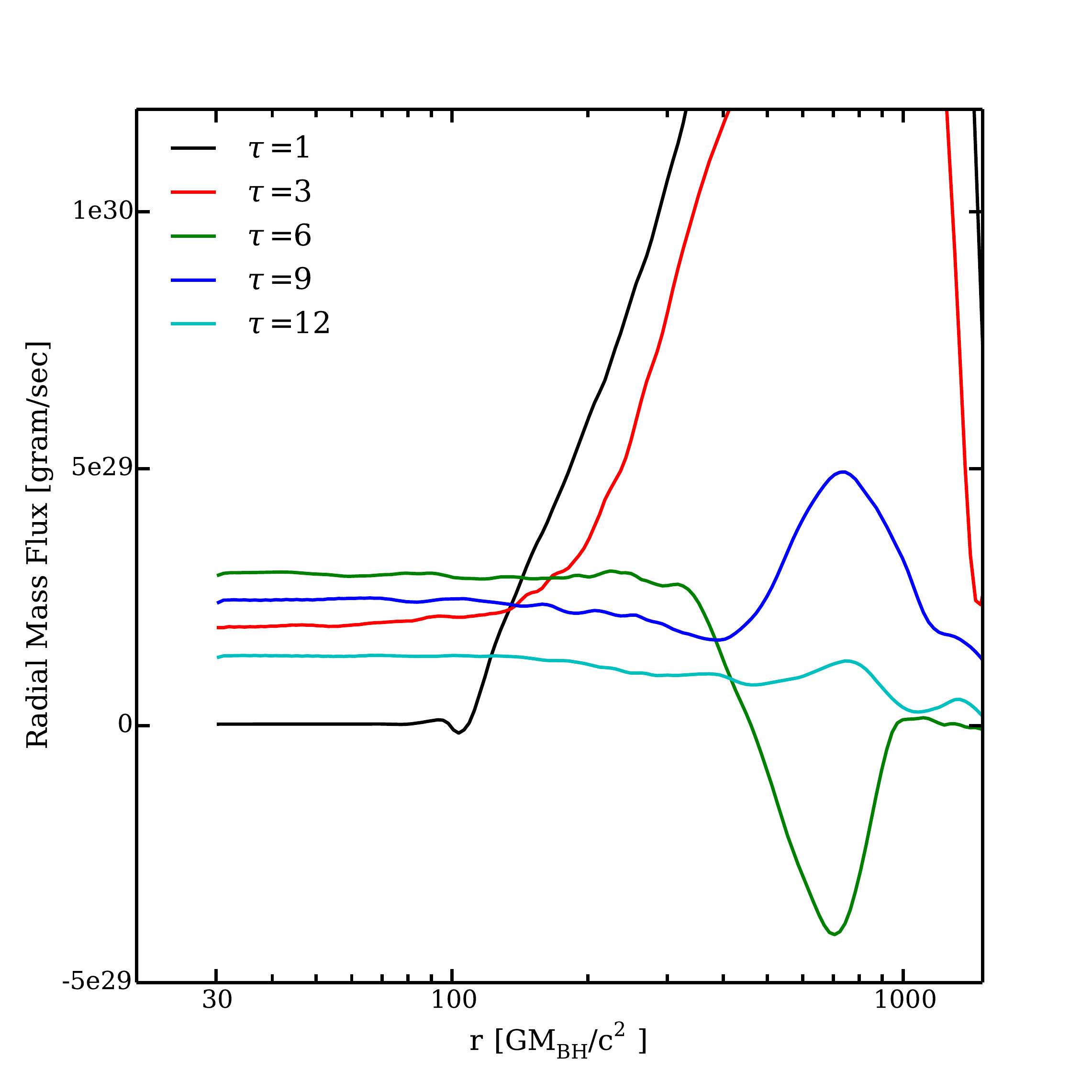}
\end{center}
\caption
{
Radial profile of radial mass flow; $F_m(r)=-\int \rho v_r rd\phi dz$.
Black, red, green, blue, and cyan colors correspond to $\tau=$1, 3, 6, 9, and 12,
respectively.   Positive mass flow is inward.    The switch in sign between $\tau=6$
and $\tau=9$ at $r \simeq 800M_{\rm BH}$ corresponds to the orbital motion
of the lump discussed in \S~\ref{sec:shock2} and also visible in Fig.~\ref{fig:mr_net}.
}
\label{r_vs_mflux}
\end{figure}

Removal of angular momentum leading to inflow well within $R_p$ can be seen in the data of
Figure~\ref{r_vs_mflux}, which shows how the radial mass flow $F_m(r)=-\int \rho v_r rdzd\phi$ depends
on distance from the black hole for various times.   Initially, there is hardly any mass inside $r \sim R_p \simeq 100M$
and there is therefore no mass flow at small radii; meanwhile, at large radii, returning streams rapidly
carry mass inward.   By $\tau = 3$, there is a steady inward flow across the range of radii $r_{\rm min} = 30M \leq
r \lesssim 125M$.    This flow is due to the inward deflection of streams at shock~1, which initially
forms at $r \sim R_p \sim 100M$.     The zone over which the inward flow extends gradually stretches outward,
as deflections at shocks~2 and 3 also begin to contribute; for $\tau \gtrsim 6$, the inflow is in an
approximate steady-state out to $r \lesssim 300M$.


We deliberately do not simulate the dynamics of mass flow at smaller radii than $30M$ because
we expect that in this region inflow is controlled by internal MHD stresses rather than the shocks
and other hydrodynamic processes treated in our simulation.   For this reason, we distinguish in
this paper between mass ``accretion" and mass ``accumulation".   We reserve the former term for
mass acquired by the black hole, having suffered enough dissipation of its orbital energy to transform
$\sim 10\%$ of its rest-mass energy into heat and possibly radiation.    Our simulation therefore treats
mass ``accumulation", the capture of mass from the disrupted star so that it resides within $\sim 100$--$1000M$
of the black hole, rather than mass accretion.

\begin{figure}
\begin{center}
\includegraphics[width=140mm]{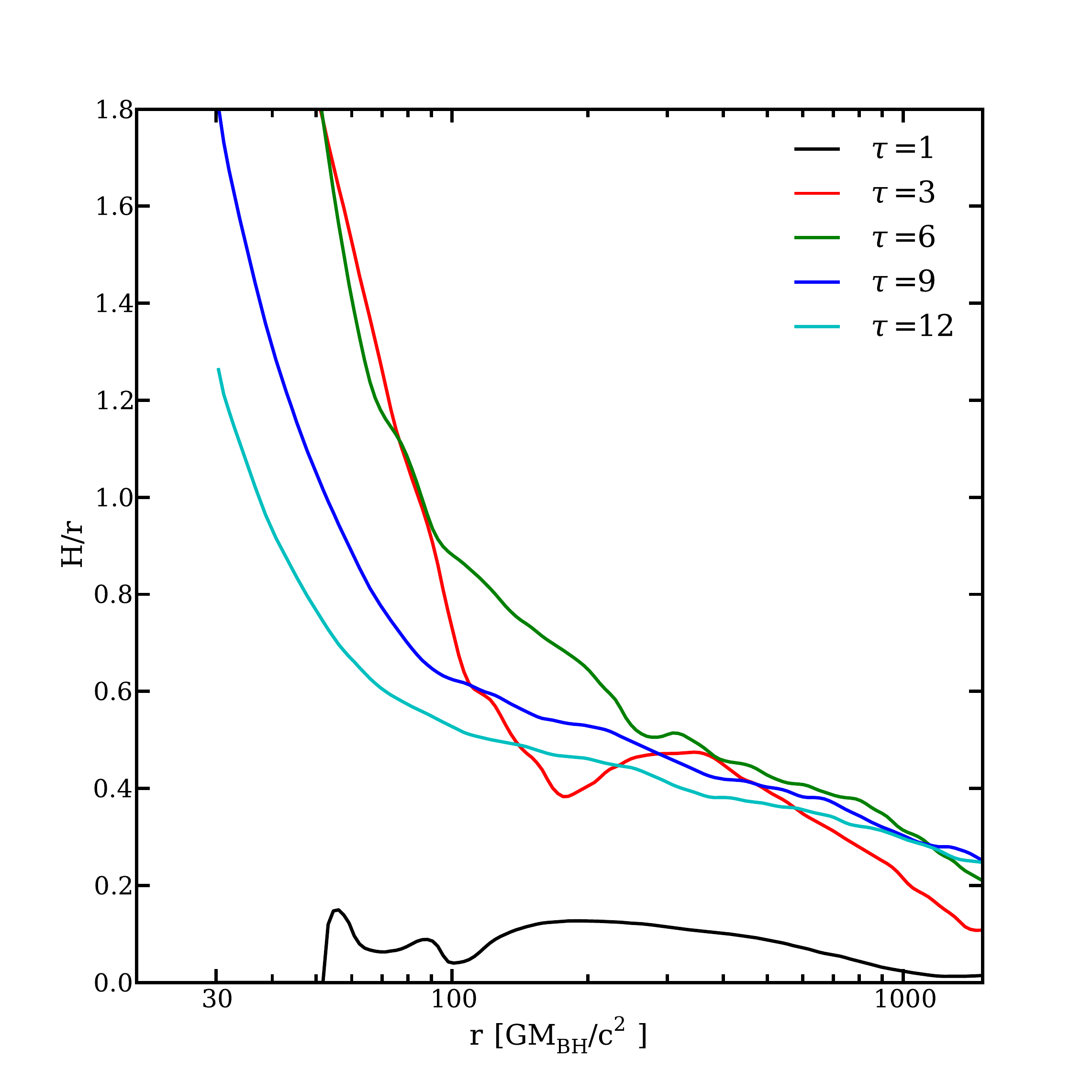}
\end{center}
\caption
{
Radial profile of azimuthally averaged disk aspect ratio $H/r$ where $H$ is scale height defined as
$\int \rho |z| dz/\int \rho dz$.
At each radius, $H/r$ is weighted by surface density and averaged in azimuthal direction.
Black, red, green, blue, and cyan colors correspond to $\tau=$1, 3, 6, 9, and 12,
respectively.
}
\label{r_vs_hor}
\end{figure}

The shocks also dissipate kinetic energy into heat: the increasing importance of pressure support causes
the geometric arrangement of the gas to grow rounder (see Fig.~\ref{fig:sigma_rho_evol}) and thicker
from the time the first tidal streams return to the black hole until $\tau \simeq 3$--6.  The vertical thickening of the
structure can be seen directly in Figure~\ref{r_vs_hor}, showing the radial profile of the azimuthally-averaged disk aspect ratio
$H/r$ ($H$ is the scale height of the disk $=\int \rho |z| dz / \int \rho dz$) at several times.    The scale height
grows rapidly from the beginning of the simulation until $\tau\sim3$, and then slowly diminishes for $\tau \gtrsim 6$.
The late-time thinning is due to the diminishing strength of the shocks combined with adiabatic cooling as the
gas spreads.    Despite the shocks' gradual weakening, the flow nonetheless remains fairly thick by conventional accretion
disk standards, with $H/r \simeq 0.5$ for $r \gtrsim 100M$ and larger at smaller radii.

In the following subsections we treat several of the flow's chief elements in greater detail.

\subsection{Shock~1 (Nozzle Shock)}

The tidal stream orbital planes have angles with respect to the midplane $\sim R_*/R_p \sim (M_{\rm BH}/M_*)^{-1/3}$.
If the streams were composed of collisionless particles, they would pass through the midplane twice, in most cases
once shortly before reaching pericenter and once more shortly after.   Because there are streams whose apocenters are both above and
below the midplane and these streams are actually fluid, this strong convergence has long been
expected to produce shocks near the pericenter \citep{Kochanek1994}.   By including high-resolution hydrodynamics, we
are able to see how pressure forces alter the structures of these shocks and prevent the streams from crossing through the
midplane.    At first, the vertical convergence of the streams merely increases the pressure of gas
near the midplane at $r \simeq R_p$.   Soon, however, this high-pressure gas begins to expand vertically;
when it does so, it intercepts new streams and a shock forms wherever newly-arriving material strikes this
high-pressure barrier (this process is analogous to the ``pancake mechanism" inside a tidally-disrupted star
described by \citet{Brassart.Luminet.2008}).

\begin{figure}
\begin{center}
\includegraphics[width=140mm]{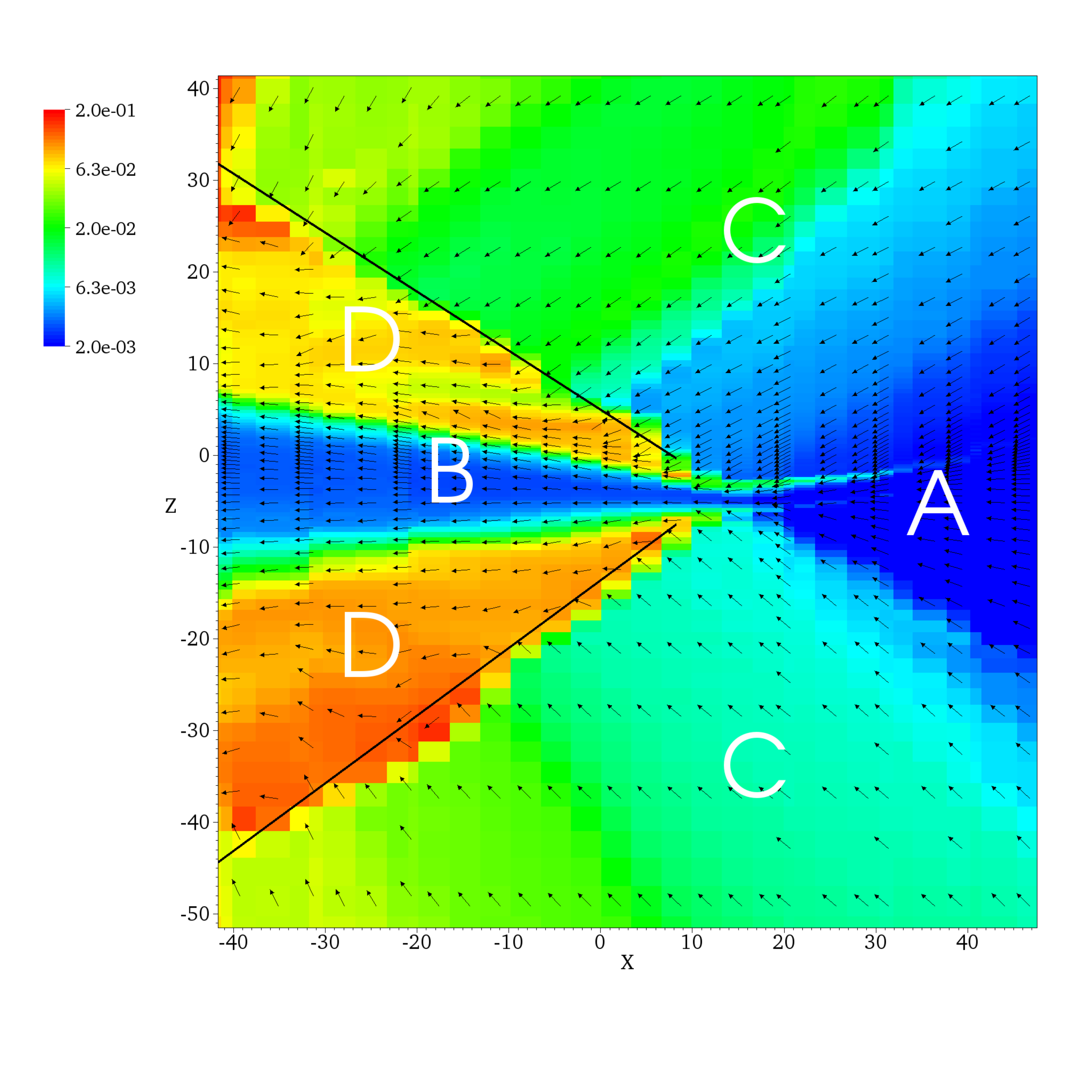}
\end{center}
\caption
{
Color contours of entropy in the X-Z plane as it cuts through shock~1 at $Y = R_p$ and $\tau = 2.35$.
Black arrows show the velocity field of the gas.
Different regions are labeled by letters and described in detail in the main text.
Shock fronts are indicated by black lines.
}
\label{nozzle_slice}
\end{figure}

The structure of shock~1 can be most clearly discerned in a plot of gas entropy (Fig.~\ref{nozzle_slice}).
Region~A in this figure is the zone of cold gas arriving at the pericenter region for the first time since it left
the star.   This gas has previously been shocked only in shock~3.   As it heads toward the midplane from both
above and below, it is compressed.    After passing through a ``nozzle" very close to the midplane, it expands on the
far side (Region~B).    Other Region~A gas, traveling farther from the midplane, encounters the wedge of high pressure
at the nozzle and is shocked, thus entering Region~D.   Gas approaching shock~1 at still greater height from
the midplane (Region~C) has higher entropy because this is not its first time through the pericenter region, and it has
already passed at least once through both shock~1 and shock~2.    It, too, passes through the shock front to enter
Region~D.

The location, size, and strength of shock~1 change over time.   In the beginning ($\tau \lesssim 1.5 $),
shock~1 stays close to the pericenter region for the more tightly-bound streams.
Later on, shock~1 stretches radially, both inward and outward as the relative contribution of
the ``Region~C" gas, the gas that has already been disturbed in either shock~1 or shock~2, grows.   The
trajectories of these streams do not go directly through the pericenter region, as the angular momentum
distribution of the gas has been broadened by previous shock passages.    Its vertical extent also changes
over time.  When it first forms, it extends only a few $M$ from the midplane; by $\tau \simeq 5$, it reaches
$\sim 50M$ above and below the midplane.  As shock~1 stretches, it also
becomes weaker: the larger the volume occupied by high pressure gas in the nozzle region, the less
incoming streams are accelerated toward the midplane by gravity and the smoother the overall flow.
The shock disappears at $\tau \simeq 7$, but immediately before it does, it spans all the way from the
inner boundary of the simulation box $r=30M$ to $r \sim 500M$. 


We close this section with a technical note.   Early in the simulation, the gas is still extremely
cold, making the vertical structure near shock~1 extremely narrow.   At $\tau \simeq 0.8$--1,
even with our most concentrated vertical cell distribution we have at some times as few
as $\simeq 5$ cells per scale height in that region.  However, the disk rapidly becomes thicker
due to shock heating (initially mostly at shock~1), and hence we do not think
this short time period of poor resolution affects the disk evolution.
After the disk is thickened enough, we deconcentrate the grid at the equator as described in \S~\ref{sec:globalgrid}.
With these adjustments, the nozzle region has $H/\Delta z \sim 40-50$ through nearly all the simulation.

\subsection{The outer shocks: shocks~2 and 3}
\label{sec:shock2}
As described in the previous subsection, shocks~2 and 3 emerge when the most bound, and therefore
earliest-returning, gas starts to collide with later-arriving gas near the apocenter of the more tightly-bound
gas's orbit.   That the velocities of the two streams are nearly perpendicular at the collision point
is a consequence of the correlation between apsidal angle and tidal stream binding energy.    In other
words, a combination of two small relativistic effects (apsidal precession of the stellar trajectory,
the exact form of the tidal stress) creates a broader spread of orbit orientations, with the contrast largely
between the earliest-returning streams and those returning later.    That early-returning streams are
oriented differently from the streams arriving soon after promotes the quick emergence of shocks~2 and 3.
Stream intersections are further promoted by additional apsidal precession when the tidal streams
return to the pericenter, but fluid effects (notably shock~1) mask that precession.
Thus, the dissipation in this shock system is a result of small, but consequential, relativistic effects in the
dynamics of tidal disruption.

\begin{figure}
\includegraphics[width=9cm]{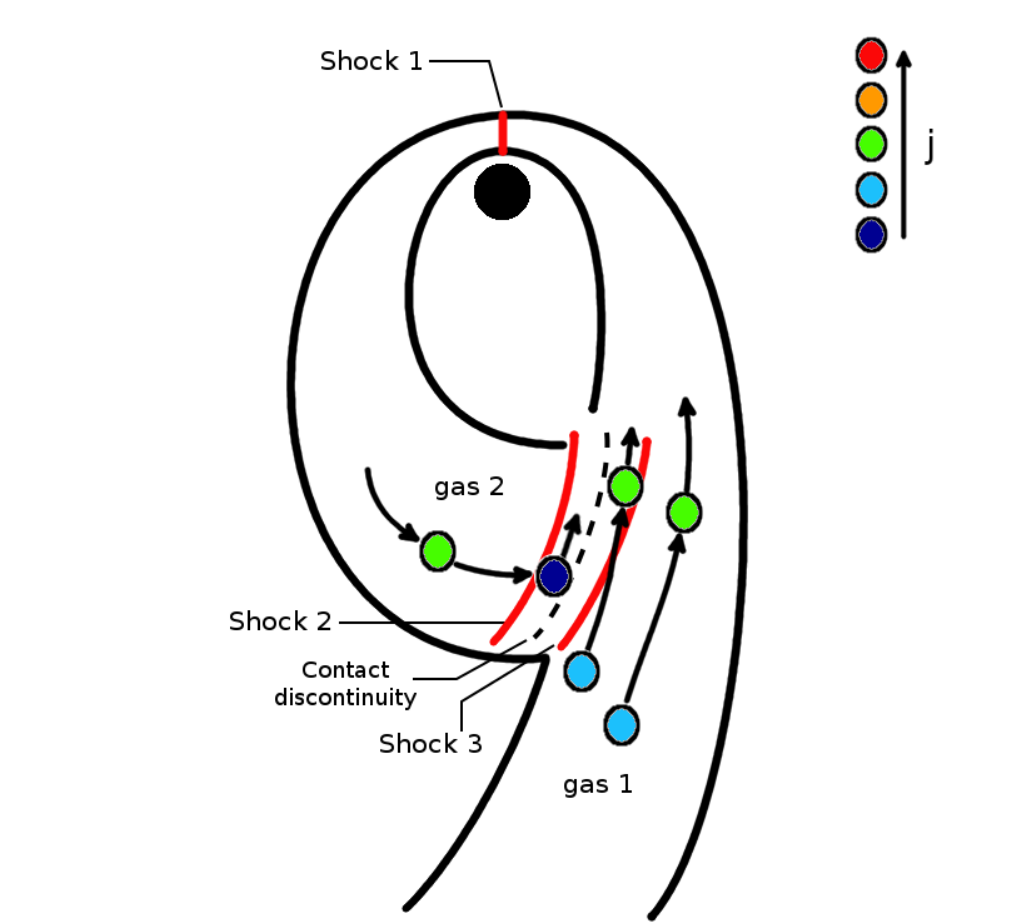}
\includegraphics[width=9cm]{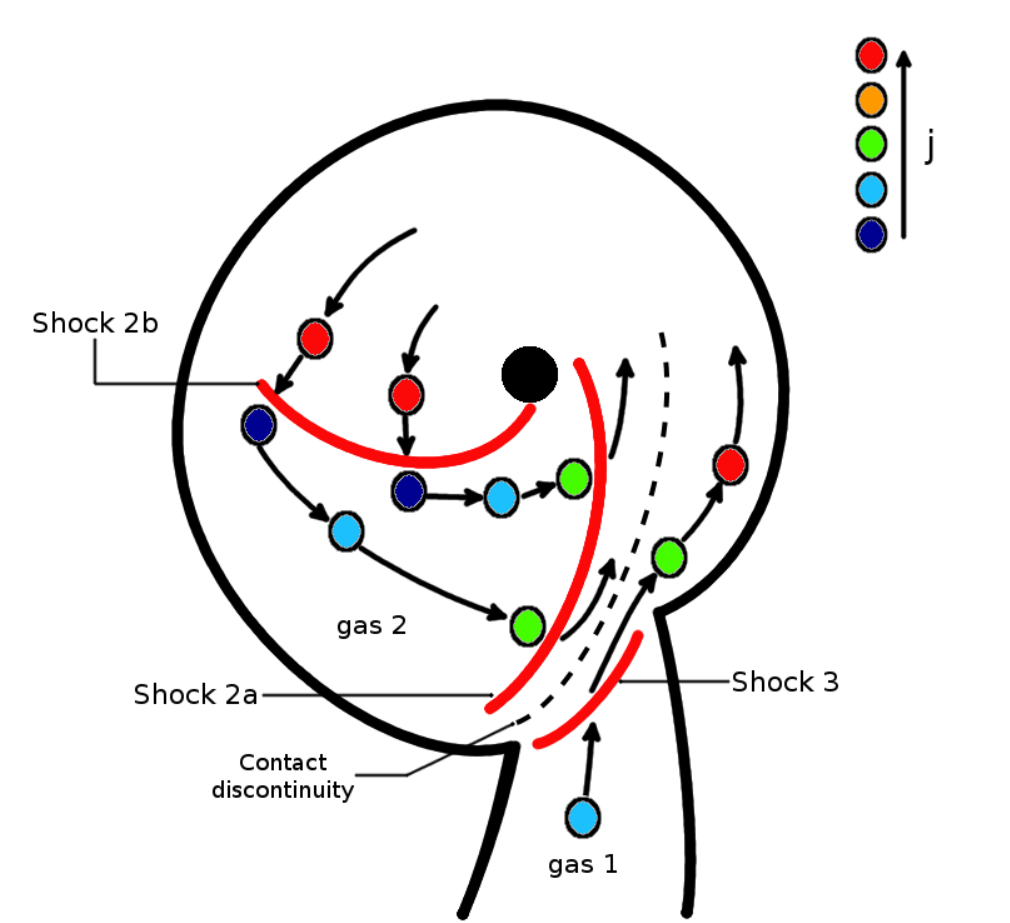}
\caption
{
Schematics of shocks~2 and 3 and the associated angular momentum redistribution at two
different times, $\tau \simeq 2$ (left) and $\tau \simeq 6$ (right).   Note that at earlier times
shock 2 has only one branch (left), whereas it isplits into two at later times (right).  Sample gas streamlines are
shown as black curves with arrows; shocks are shown in red.   The large black disk is the black hole; the thick
black curves without arrows are rough indications of the boundaries of the flow.   Relative angular
momentum is color-coded as shown.  The contact discontinuity between shocked gas-1 and gas-2
is shown by a dashed black curve.  Separations are not drawn to scale in order to emphasize the
sequence of events.
}
\label{schematic}
\end{figure}

To understand the structure, evolution, and consequences of these shocks, we have constructed
a schematic shown in Figure~\ref{schematic}.    In this diagram, we designate the newly-arriving
stream gas-1 and the stream that has already swung through the pericenter region gas-2.
The first step in the shock formation process is a compression of the material in gas-2 as it decelerates
while climbing out of the black hole's potential well toward apocenter; this is a purely kinematic effect and
would take place in a collisionless gas in the same way as in a fluid.    The deceleration creates
a density enhancement just prior to the point of intersection between gas-1 and gas-2.

Shocks~2 and 3 form when gas-2 strikes gas-1 from the side.   At shock~3, gas-1 is shocked, raising
its pressure.   This high pressure zone creates shock~2, which shocks gas-2.    Shocked gas-1 and gas-2 then
have the same pressure, but are separated by a contact discontinuity (left panel of Fig.~\ref{schematic}).
Because the density of gas-2 was already elevated by its deceleration, its compression in the shock creates
a region of even higher density.    As time goes on, the mass contained in this feature grows, widening the high
pressure region between shock~2 and the contact discontinuity, and moving the location of the shock front farther
upstream.    When gas-2 passes through this front, it is deflected in a way that causes it to move more
radially inward; as it does so, it is accelerated by gravity, and its pressure falls by adiabatic expansion.
When the separation between the shock front and the contact discontinuity grows large enough, the
shocked gas loses enough pressure by the time it approaches the contact discontinuity that it shocks for
a second time (right panel of Fig.~\ref{schematic}).   From this point onward, shock~2 is divided
into two branches, which we call 2b (the one farther upstream) and 2a (the one close to shock~3).

The first passage through shock~2 of the mass associated with the peak in the mass-return rate leads to
a surprising effect: the density concentration becomes a quasi-coherent object
and traces an eccentric orbit around the black hole (its radial oscillation is visible in Fig.~\ref{fig:mr_net}).
When this ``lump" reaches apocenter, the position of shock~3 moves outward; when it is near pericenter,
shock~3 moves closer to the black hole.   

\begin{figure}
\begin{center}
\includegraphics[width=140mm]{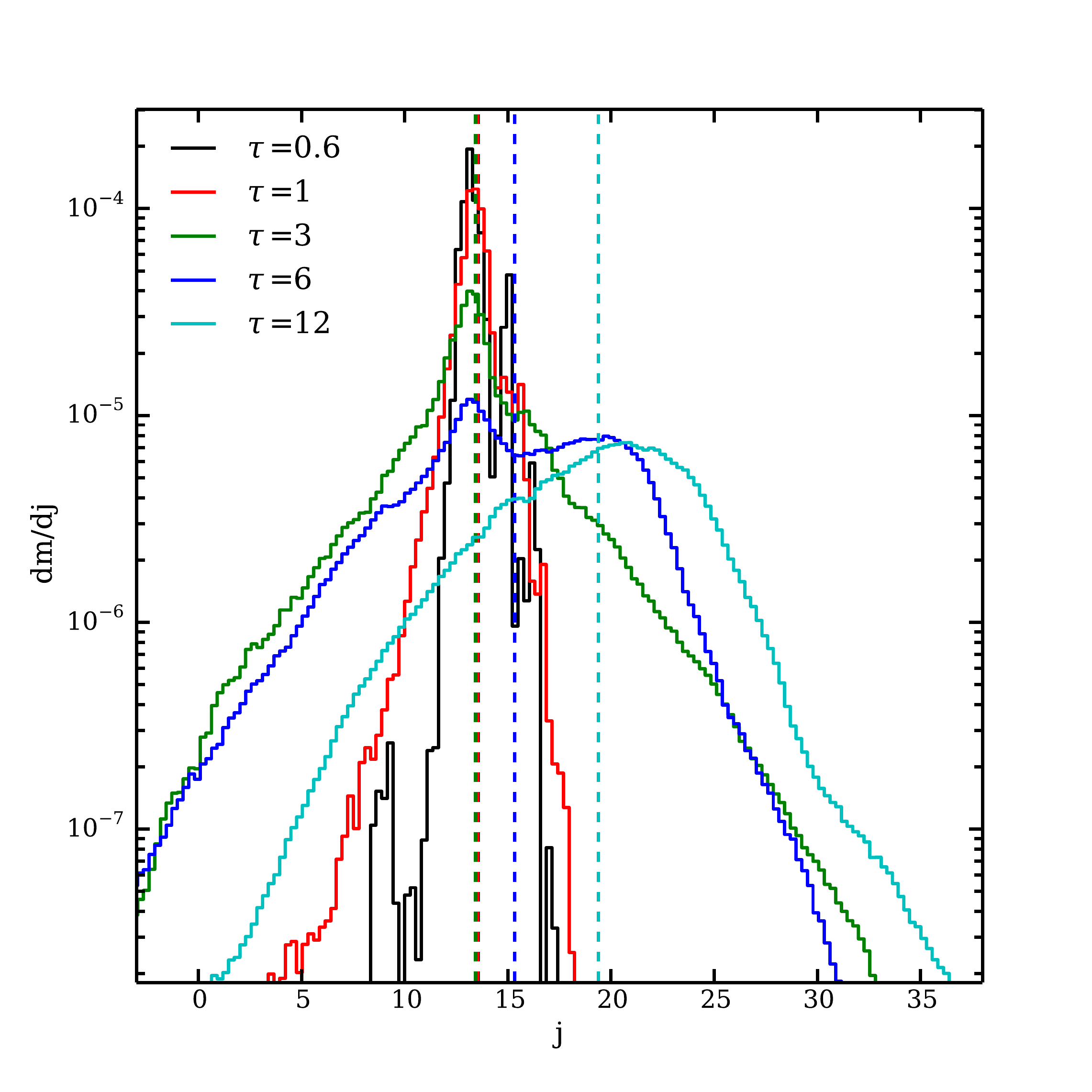}
\end{center}
\caption
{
Mass distribution of specific angular momentum in the global simulation.
Different lines represent the distribution at different times: $\tau=0.6$ (black), 1 (red), 3 (green),
6 (blue), and 12 (cyan).
Broken lines show the average value over the distribution of corresponding color.
$dm/dj$ is dimensionless because $m$ is in units of $M_{\rm BH}$ while $j$ is in
units of $(r/R_g)(v_\phi/c)$.
}
\label{distribution_j}
\end{figure}

Angular momentum redistribution is accomplished by deflections at the shocks (\cite{Rosswog.et.al.2009} also report
a shock that appears to resemble shock~2 in its early stages and remark briefly that it redistributes angular momentum).
The shock velocity of shock~2, both before and after its split, is directed roughly azimuthally and against the sense
of the gas orbits; it therefore reduces the angular momentum of gas-2 and deflects it inward.   The angular momentum
lost by gas-2 is transferred to gas-1, deflecting it outward.    As a corollary, gas-2 also does work on gas-1, reducing
its own orbital energy while increasing that of gas-1.   However, this picture becomes a bit more
complicated when shock~2 splits into two separate shocks.    During this period, angular momentum
is transferred from incoming gas-2 at smaller radii to gas-2 shocked by shock~2b at larger radii that
has then turned inward.   Only after a second deflection at shock~2a does the angular momentum and energy
reach gas-1.    Of course, gas-1 becomes gas-2 after taking a turn through the pericenter region, so the
net result of this process is to broaden the distribution of specific angular momentum.  Figure \ref{distribution_j}
shows how the box-integrated angular momentum distribution changes over time as a result of these processes. 
From its initial very narrow distribution, the spread in angular momentum grows substantially by $\tau = 3$.
At later times, the growth in width slows, but, as more mass at the low end of the angular momentum distribution falls through
the inner boundary, the average value of the specific angular momentum (vertical lines in the figure) shifts
toward higher values.   The energy distribution also widens, particularly toward more negative orbital energy,
but because it begins far broader than the angular momentum distribution, the effect is less noteworthy.

\subsection{Heating}

To determine the heating rates of the different shocks, we compute the divergence of the heat flux carried by the fluid
and integrate over a volume containing the shock in question.   This method suffers from certain inaccuracies (mixing
heating by adiabatic compression with true entropy generation, by-eye choice of the specific volumes for integration), but
in these circumstances appears to provide the best feasible option.   

\begin{figure}
\begin{center}
\includegraphics[width=140mm]{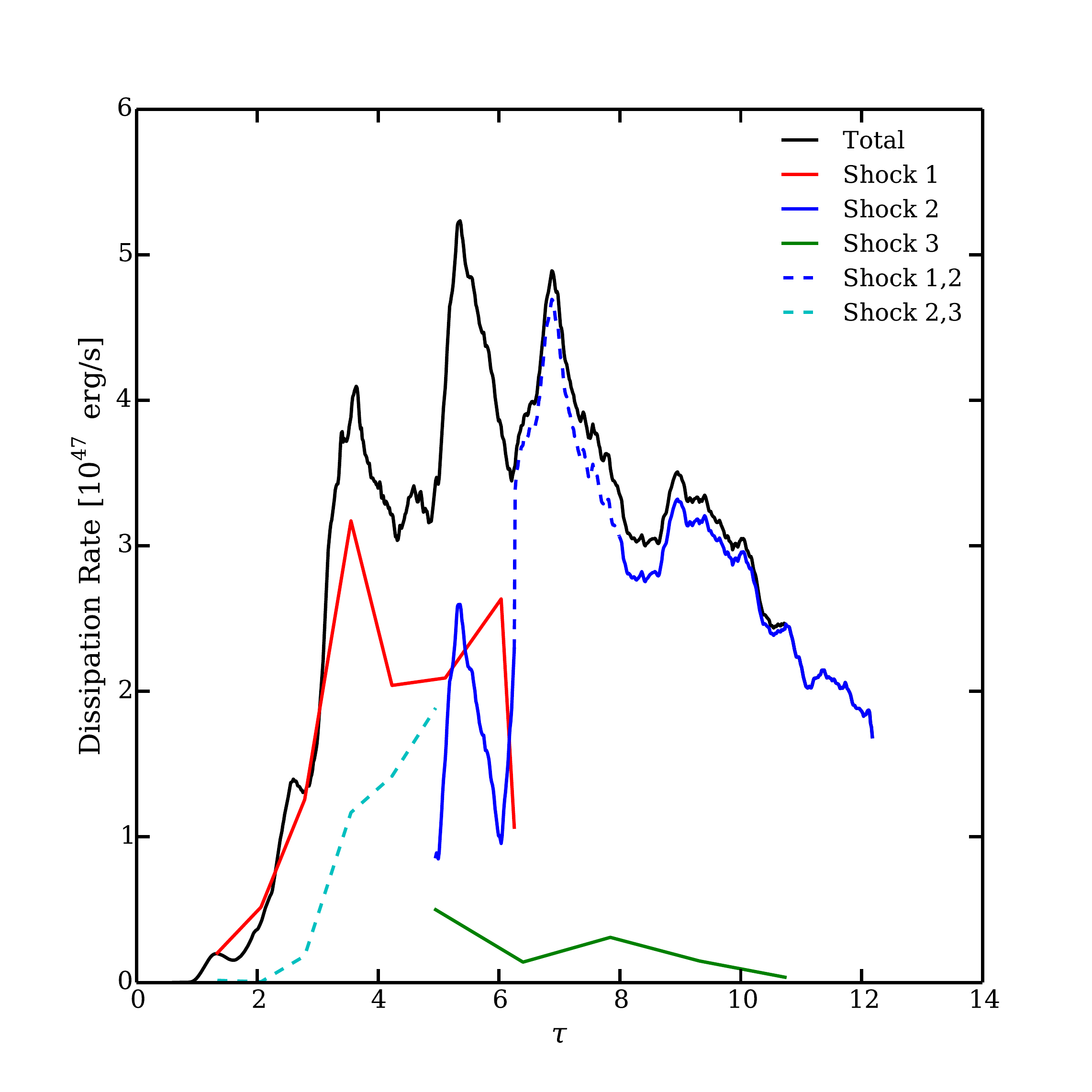}
\end{center}
\caption
{
Heating rate as a function of time.    Total (black curve); shock~1 (red solid curve); shock(s)~2 (blue solid curve);
shock~3 (solid green curve); shocks~2 and 3 together (dashed cyan curve); shocks~1 and 2 together (dashed blue curve).
}
\label{dissipation_rate}
\end{figure}

When we apply this method to our data, we find that different shocks create the most heat at different times
(Fig.~\ref{dissipation_rate}).   Shock~1 forms first, so it dominates the heating early on.    Once shocks~2 and 3 emerge,
their heating rate grows.    Shock~3 is a strong shock, with Mach number $\sim 10$, and heats the gas from
very low temperatures to $\sim 0.1\times$ the local virial temperature.
Shock~2 has a lower Mach number because its upstream gas has already been heated in shock~1, but  nonetheless
produces comparable heating.   Unfortunately, until $\tau \sim 5$, shocks~2 and 3 are so close together we cannot
reliably separate the volumes affected by them in order to compute their heating rates independently.   Nonetheless,
it is possible to say that by $\tau \simeq 3$--4, the heating in shocks~2 and 3 combined grows to about half that in shock~1,
and is almost as large as the heating in shock~1 by $\tau \simeq 5$. 

During this entire earlier period ($\tau \lesssim 5$), the total heating rate grows, increasing by a factor $\simeq 4$
between $\tau \simeq 2$ and $\tau \simeq 5$.    It remains at about the same level ($\sim 4 \times 10^{47}$~erg/s
for the parameters of our simulation) until $\tau > 8$.   However, the dominant contributors change after
$\tau \simeq 5$.   Heating by shock~1 falls sharply because this is the time at which its Mach number becomes
$\sim 1$.   Meanwhile, shocks~2  extend to smaller radius,
where the larger orbital speeds permit their heating rates to grow.   This extension also leads to such close proximity
to the dying shock~1 that we cannot cleanly separate the heating rate in shock~1 from that in shocks~2 from
$\tau \simeq 6$ until $\tau \simeq 8$.    During the period after $\tau \simeq 3$, the rate at
which newly-arriving matter reaches shock~3 diminishes, while the mass flux passing through the other shocks
does not decrease nearly as much because matter passes through them repeatedly.   Consequently, the
heating in shock~3, although significant earlier, becomes negligible after $\tau \simeq 5$.  Thus, from $\tau \simeq 8$
onward, the heating is primarily associated with shocks~2.

\begin{figure}
\begin{center}
\includegraphics[width=140mm]{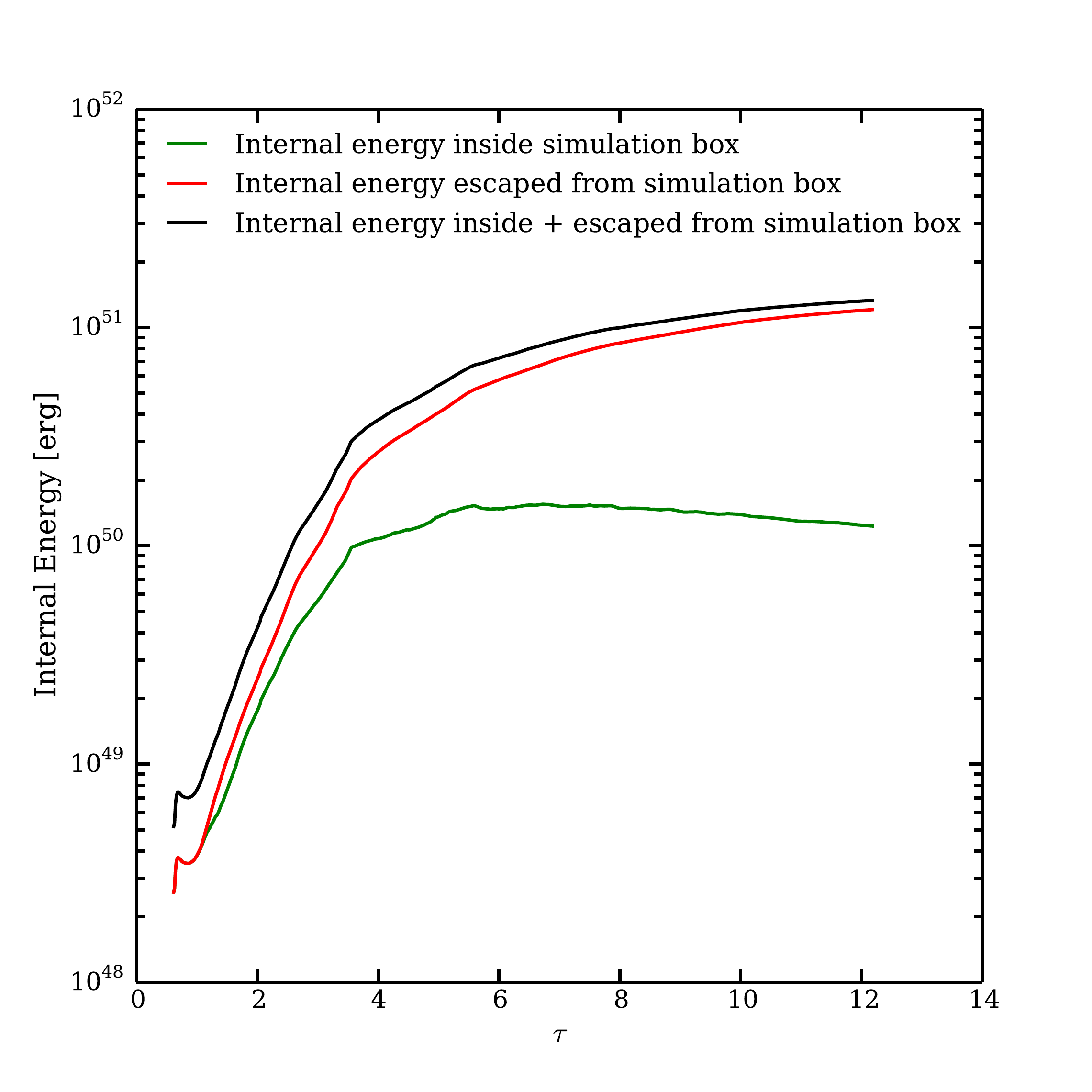}
\end{center}
\caption
{
Time evolution of total internal energy inside simulation box (green), total internal energy escaped
from the simulation box (red), and sum of them (black).
}
\label{t_vs_utot}
\end{figure}

As a result of all this heating,  the total internal energy grows rapidly until $\tau \simeq 5$; afterward, it
is roughly constant (Figure~\ref{t_vs_utot}).   This is because most of the shock heating takes place at
small enough radii that new heat can be drained through the inner boundary by the radial inflow.

Although the total internal energy generated by the end of the simulation is $\sim300$ times more than the
initial internal energy, it still is only $\sim0.1$\% of the rest-mass energy of the star because the shocks responsible
for the heating occur predominantly at large radius.    For gas to settle near
$R_p$, it would need to lose an order of magnitude more heat.   In this sense, the hydrodynamic
processes taking place in this simulation do not solve the dissipation problem posed in the Introduction.   It
also follows immediately that the heating we have studied can account for at most only a very small fraction
of the total energy released in a tidal disruption event if ultimate accretion onto the black hole has the usual
relativistic radiative efficiency $\sim 10\%$ in rest-mass units.

\subsection{Circularization}
\label{sec:circularization}

Classical accretion disks are, in most instances, imagined as very thin structures in which
the material travels on very nearly circular orbits while moving inward on a timescale much
longer than an orbital period.   As a result, departures from axisymmetry are quite small.   In
a tidal disruption event, it is an interesting question whether, and to what degree, such a state
is achieved.   To measure progress toward such ``circularization", we define several independent
measures.

\begin{figure}
\begin{center}
\includegraphics[width=140mm]{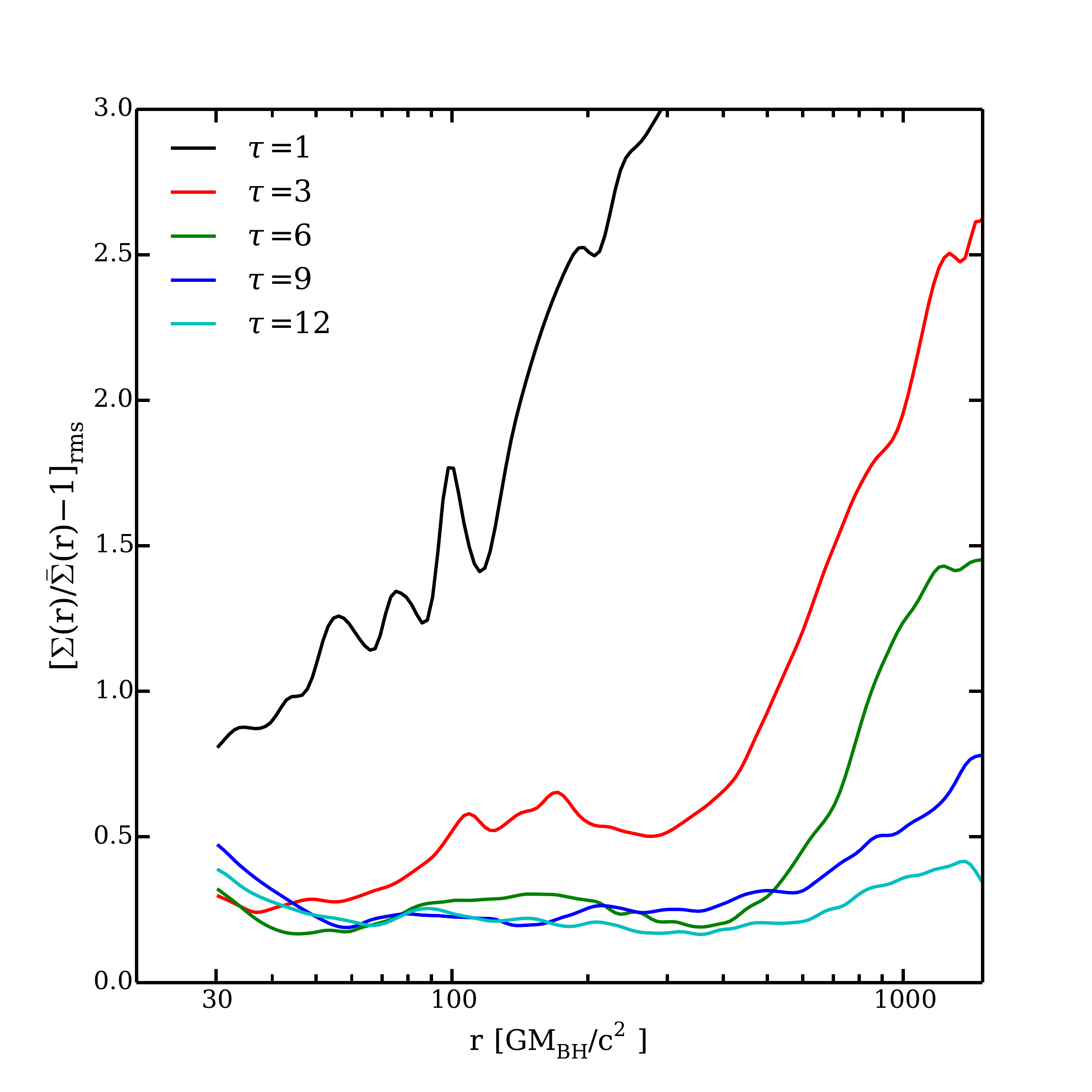}
\end{center}
\caption
{
Radial profile of {\it rms} azimuthal fractional surface density fluctuation,
$\delta(r)=[\Sigma(r,\phi)/{\bar \Sigma}(r) - 1]_{\rm rms}$ at several times.}
\label{r_vs_rms_sigma}
\end{figure}

The first is the level of nonaxisymmetry in its surface density.   Figure~\ref{r_vs_rms_sigma} shows the
{\it rms} fractional surface density fluctuation around the azimuthal coordinate at each radius,
$\delta(r) \equiv [\Sigma(r,\phi)/{\bar\Sigma}(r)-1]_{\rm rms}$, at several times.     Here ${\bar\Sigma}$
is the mean over azimuthal angle $\phi$.   At large radii ($r \gtrsim 400M$--$600M$),
$\delta$ remains $ \sim 1$ until very nearly the end of the simulation; this large asymmetry reflects the narrowness of the
tidal streams.   The large asymmetry at all radii when $\tau = 1$ is due to the same cause.
Closer in ($100M \lesssim r \lesssim 400M$), the disk remains quite asymmetric ($\delta \simeq 0.5$)
until $\tau \gtrsim 4$, when $\delta$ falls to $\simeq 0.2$--0.3 and stays at that level until the
end of the simulation.   In this region, the asymmetries are largely associated with shocks 1, 2a, and 2b.
At still smaller radii ($30M \leq r \lesssim 50M$), there is a modest increase in asymmetry that
arises from the inward funneling of particularly low angular momentum gas.     The upshot from this
measure is that the flow is quite far from circular at least until $\tau \simeq 4$, and remains significantly
non-circular even at $\tau \simeq 12$.

\begin{figure}
\begin{center}
\includegraphics[width=140mm]{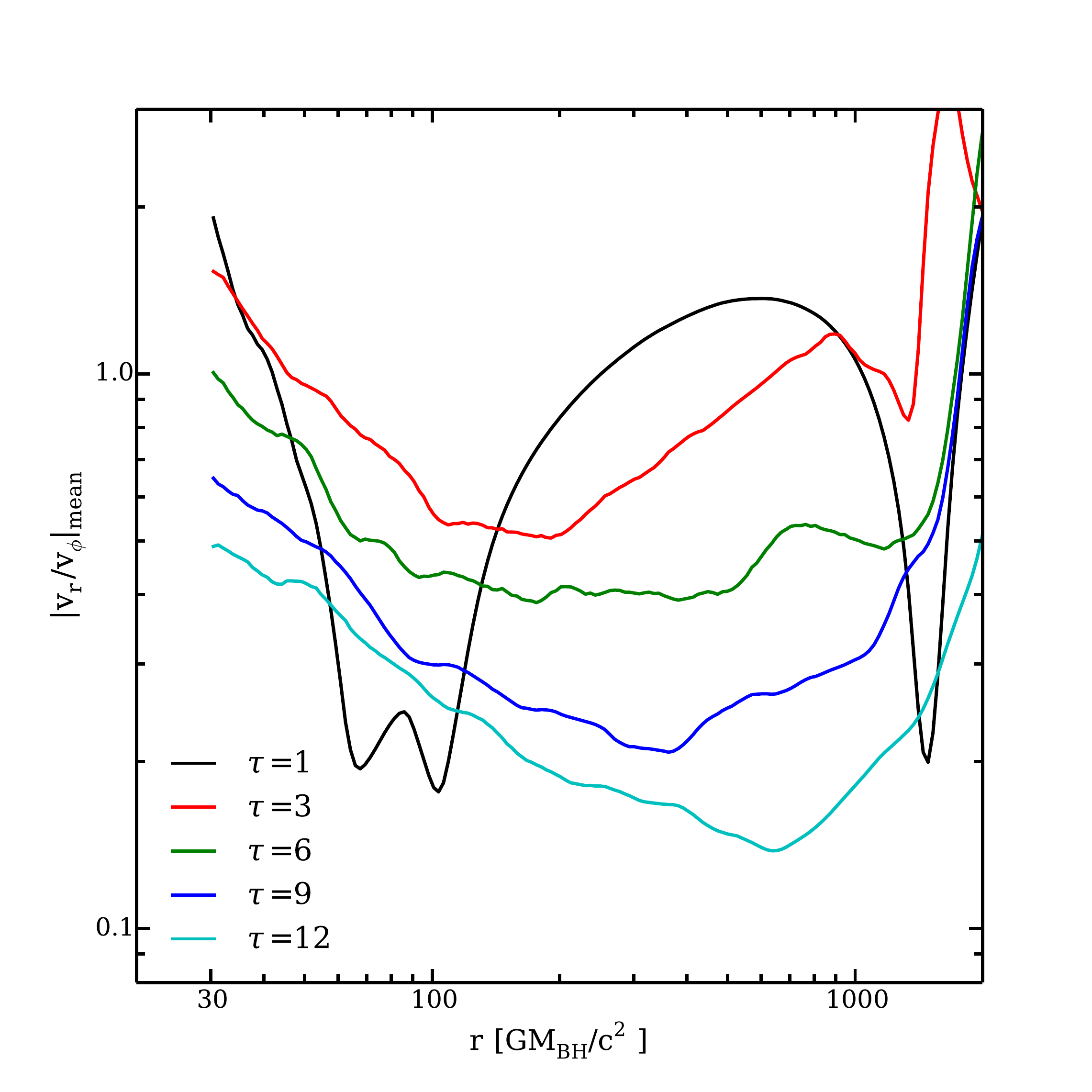}
\end{center}
\caption
{
Radial profile of the mass-weighted azimuthal average of the ratio of radial to azimuthal velocity.
Black, red, green, blue, and cyan colors correspond to times $\tau=$1, 3, 6, 9, and 12,
respectively.
}
\label{vr_o_vph}
\end{figure}

Our second measure is the ratio of radial to azimuthal velocity $|v_r/v_\phi|$.
Whereas $\delta (r)$ measures morphological circularity, $|v_r/v_\phi|$ measures the circularity
of orbital motions.  Figure~\ref{vr_o_vph} shows the radial profile of the density-weighted average of $|v_r/v_\phi|$,
\begin{equation}
R_v(r) = |v_r/v_\phi|_{mean} = \frac{\int \rho|v_r/v_\phi| rd\phi dz}{\int \rho rd\phi dz}.
\end{equation}
As can be seen in this figure, the degree of orbital circularity improves steadily over time.   Early
on ($\tau \simeq 1$), $R_v$ has a rather complex dependence on radius; this dependence is
due to the initial tidal stream structure.    From $\tau \simeq 2$ onward, $R_v$ declines steadily
at all radii from $\sim 1$ to $\simeq 0.2$--0.4 by the end of the simulation.    Just as steadily,
the radius at which $R_v$ is minimum moves outward, reflecting the progressively larger
portion of the flow that is at least partially circularized.   In fact, judged on an
absolute standard, the fluid orbits are approaching a reasonable degree of circularity by
late times.    The relative magnitude of the radial velocity would also be roughly in line
with expectations for the mean inflow speed if a disk this thick were subject to the MHD
stresses that ordinarily drive accretion, even though these stresses are absent here.

\begin{figure}
\begin{center}
\includegraphics[width=140mm]{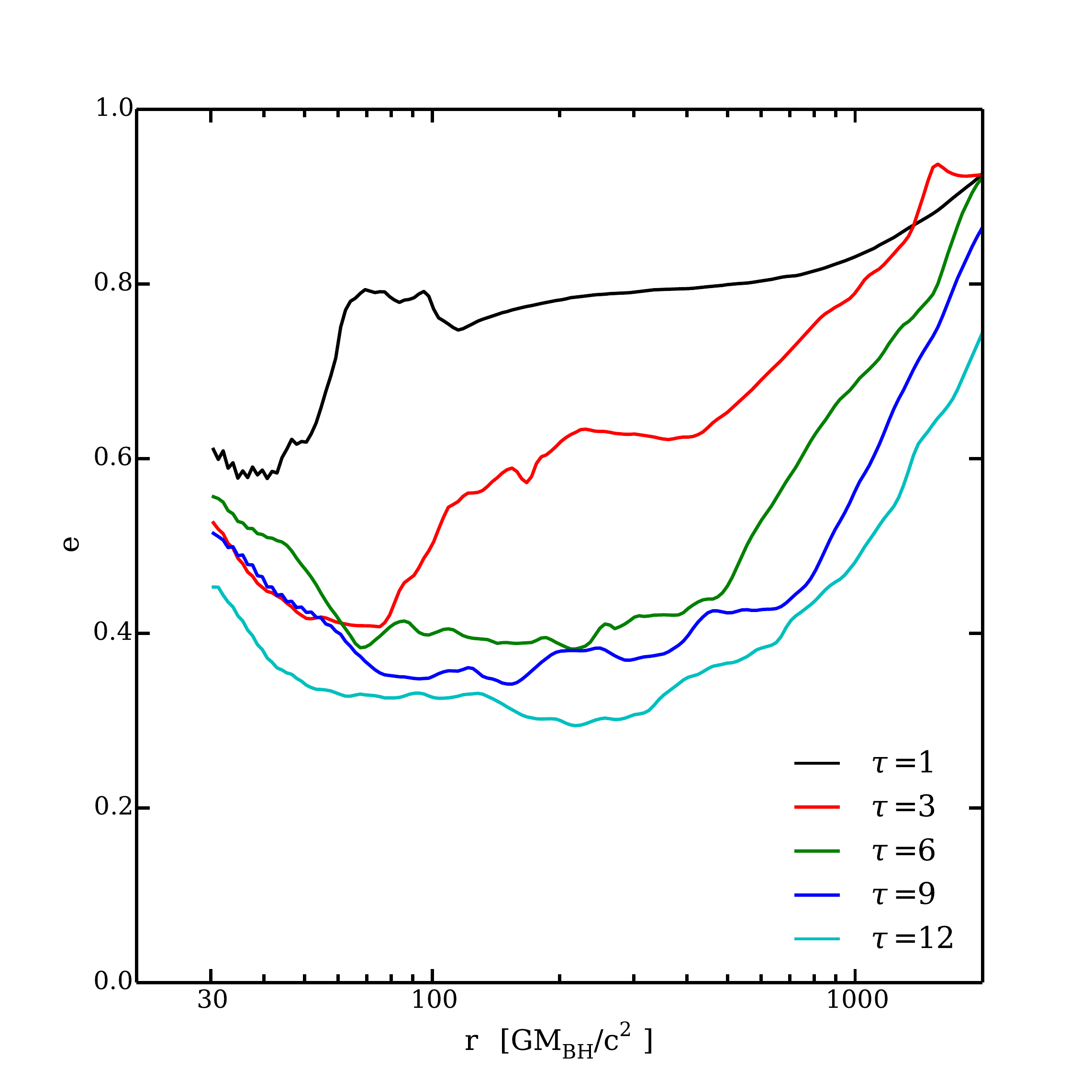}
\end{center}
\caption
{
Radial profile of mass-weighted mean eccentricity.
Black, red, green, blue, and cyan colors correspond to $\tau=$1, 3, 6, 9, and 12,
respectively.
}
\label{r_vs_ecc}
\end{figure}

A third measure is given by the nominal eccentricity of the fluids' orbits; that is, the eccentricity
of a test-particle orbit with the same specific angular momentum and orbital energy:
\begin{equation}
e = \left[1 + 2\varepsilon j^2/(GM)^2)\right]^{1/2},
\end{equation}
where the specific orbital energy $\varepsilon \equiv (1/2)v^2 - GM/r$ and the specific angular
momentum $j = r v_\phi$.    For our diagnostic, we construct its mass-weighted azimuthal and vertical average
\begin{equation}
e_r(r) = \frac{\int \rho e rd\phi dz}{\int \rho rd\phi dz}.
\end{equation}
Figure \ref{r_vs_ecc} shows the radial profile of $e_r$ for several times ranging from the very
beginning of the simulation to the end.     This measure also decreases steadily over the
duration of the simulation, although its rate of decrease decelerates after $\tau \simeq 6$.
By the end of the simulation, the mean eccentricity has diminished to $\simeq 0.3$--0.4
over most of the flow's volume.   In other words, this diagnostic shows a level of circularity
similar to the one implied by $R_v$.

\begin{figure}
\begin{center}
\includegraphics[width=140mm]{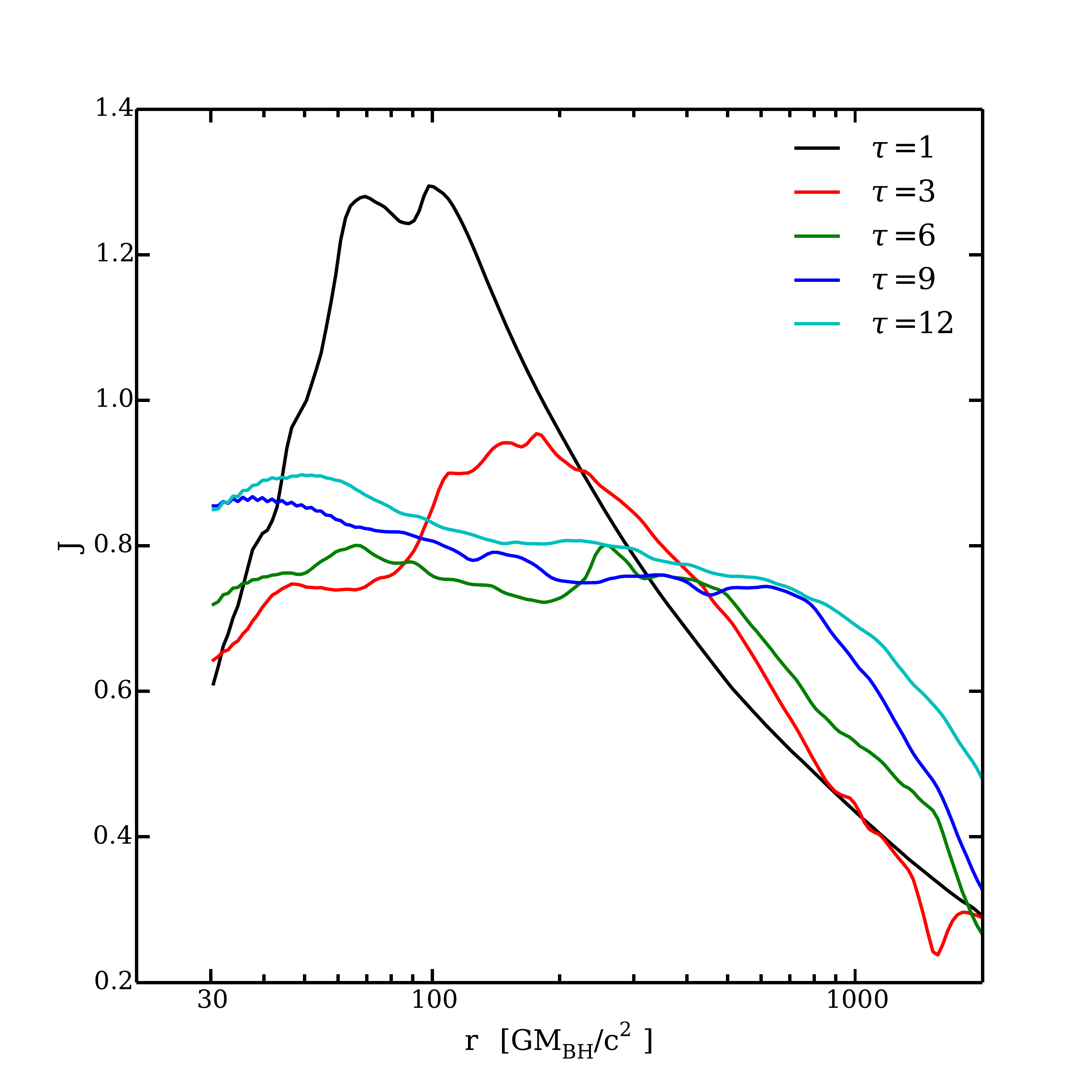}
\end{center}
\caption
{
Radial profile of density weighted average of $J$, the ratio of the mass-weighted specific
angular momentum to that required for a circular orbit, at several times.
Black, red, green, blue, and cyan colors correspond to $\tau=$1, 3, 6, 9, and 12,
respectively.
}
\label{r_vs_j_o_jkep}
\end{figure}

Lastly, we examine the proportion of rotational support against gravity as measured by the ratio between
the gas's specific angular momentum and that required for a circular orbit with that fluid element's instantaneous
radius; in a classical disk, this ratio would be unity.    Figure~\ref{r_vs_j_o_jkep} shows the mass-weighted average
of this quantity as a function of time at several radii.
\begin{equation}
J(r) = j_{mean}(r)/j_{K}(r) = \frac{\int \rho j rd\phi dz}{j_K(r)\int \rho r d\phi dz}.
\end{equation}
Initially, almost complete rotational support is implied at radii near and inside $r_p$ ($60M \lesssim r \lesssim 100M$),
while rotational support plays a much smaller role everywhere else.   However, over time all locations
converge toward a value of $J \simeq 0.8$, a bit greater for $r < 100M$, somewhat less at $r > 1000M$.
This much angular momentum is enough to create a somewhat flattened structure, but $1 - J$ is large enough
that substantial pressure support is also implied.

\subsection{Actual mass accumulation rate}

As discussed in Sec.~\ref{sec:intro}, it has long been assumed that in tidal disruption events the ballistic
mass-return rate (after a quick rise to a maximum at $\tau \simeq 1$) is well approximated by $(M_*/3) \tau^{-5/3}$,
and that this same expression is a good predictor of both the rate
$\dot M_{\rm disk}$ at which matter accumulates in an accretion disk of radius $\simeq 2R_p$ and the rate $\dot M_{\rm acc}$
at which matter enters the central black hole.    We have already shown (Sec.~\ref{sec:mass_return}) that the
ballistic mass-return rate reaches its maximum at $\tau \simeq 1.5$ and asymptotes to a curve $\propto \tau^{-5/3}$,
but its maximum value is about twice the value predicted by the $\tau^{-5/3}$ curve at the time of that maximum
and asymptotes to $\tau^{-5/3}$ only for $\tau \gtrsim 3$.

\begin{figure}
\begin{center}
\includegraphics[width=140mm]{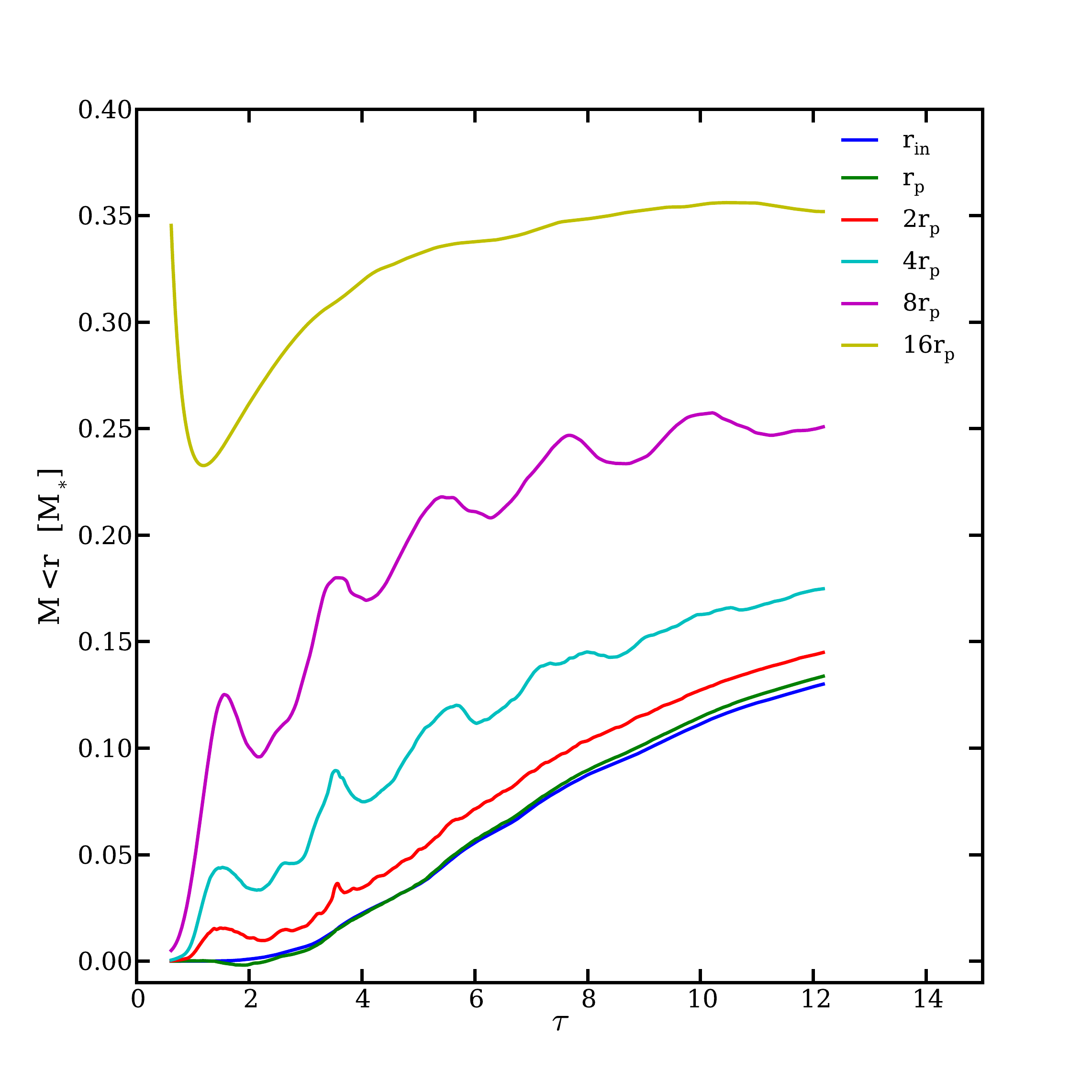}
\end{center}
\caption
{
Mass interior to a given radius as a function of time.   The various radii plotted are: $16R_p$ (yellow curve),
$8R_p$ (magenta curve), $4R_p$ (cyan curve), $2R_p$ (red curve), $R_p$ (green curve), $r_{\rm in}$ (blue curve).
}
\label{fig:mr_net}
\end{figure}

Our hydrodynamic results indicate stronger contrasts with the usual assumptions about $\dot M_{\rm disk}$
and $\dot M_{\rm acc}$.    First of all, the ``disk" structure forms at considerably larger radius than $\simeq 2R_p$.
As shown in Figure~\ref{fig:mr_net}, even at $\tau \simeq 12$, only $\simeq 1/3$ of the bound mass has moved
within $2R_p$ of the black hole.    In fact, to reach the point within which half the bound mass can be found
requires moving out to $r \simeq 6R_p$, and only $\simeq 3/4$ of the bound mass is within $16R_p$ at $\tau = 12$.    Thus,
most of the mass accumulates at scales comparable to the semi-major axis of the most bound material $a_{\rm min}$
(in this case, $\simeq 500M = 5 R_p$), rather than near $r \sim R_p$.

\begin{figure}
\begin{center}
\includegraphics[width=140mm]{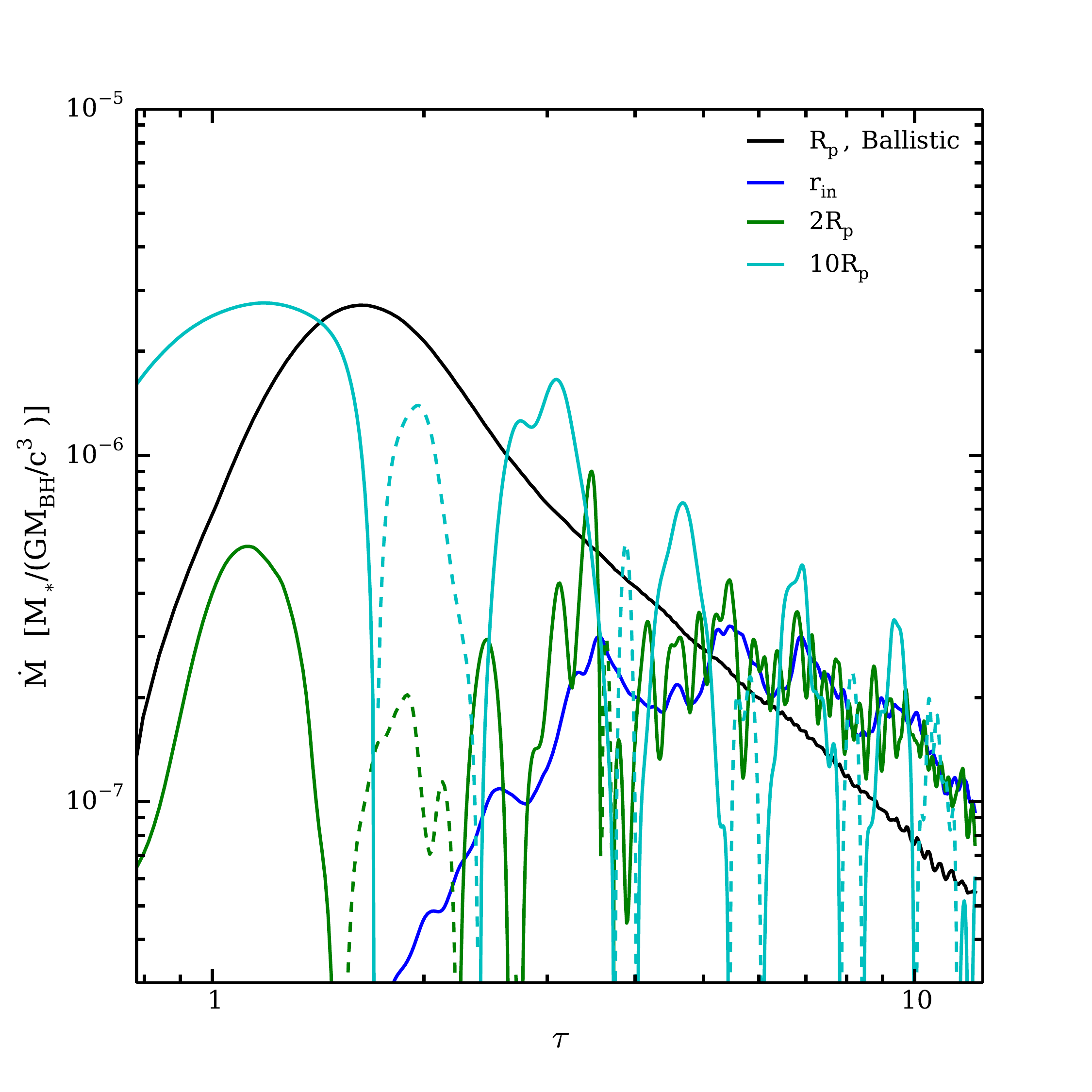}
\end{center}
\caption
{
Several different measures of the mass inflow rate: ballistic, i.e., pseudo-particles (black curve); the mass flow
through the inner boundary ($\dot M_{\rm acc}$, blue curve); the rate of increase of the mass within
$2R_p$ (green curve); and the rate of increase of the mass within $10R_p$ (cyan curve).   There are
frequent episodes of net {\it outflow}; these are denoted by dashed curves whose color corresponds to
the radius of reference.
}
\label{fig:mdot}
\end{figure}

The extended mass distribution and its azimuthal asymmetry together make the accumulation of mass in the
accretion flow considerably slower than the mass-return rate would indicate.   Half the ultimate mass within $2R_p$ arrives
after $\tau \simeq 5$; even bringing half the ultimate mass within $8R_p$ isn't accomplished
until $\tau \simeq 3$.   Second, and more surprisingly, $M(<r)$ oscillates as a function of time, particularly for
$2R_p \lesssim r \lesssim 10R_p$, i.e., $\dot M_{\rm disk}$ changes sign frequently across that entire range
of radii (Fig.~\ref{fig:mdot}).    At early times, a great deal of mass moves within $10R_p$ as the tidal streams
begin their approach; however, after passing near pericenter, the matter in those streams swing back out, reaching
as far as $10R_p$ (see also the first panel of Fig.~\ref{fig:sigma_rho_evol}).  This effect leads to an oscillation in
the mass within $\simeq 8R_p$ because the streams are not evenly spread in orbital phase---there is a concentration
in the lump mentioned earlier.    Thus, any measure of $\dot M_{\rm disk}$ follows the ballistic return-rate curve only with
sizable fluctuations, both above and below the ballistic return-rate, until $\tau \gtrsim 5$.

Stream deflection of shocks does, however, partially aid in directing mass inward, supplementing the usual source of internal
stress in accretion flows, correlated MHD turbulence.   However, the rate
at which this mechanism drives matter through our inward boundary is well below the ballistic return
rate until $\tau \gtrsim 5$, and the rate reached at that time is only $\sim 0.1 \times$ the peak rate of
mass-return as predicted by a purely ballistic model.    This rate, of course, is still not the rate $\dot M_{\rm acc}$
at which mass will flow into the black hole; that rate depends on the strength of further angular momentum transport,
presumably due to MHD turbulence.

These results contrast strongly with classical expectations.   By the end of the simulation at $\tau \simeq 12$,
28\% of the bound mass has passed through the inner boundary.     In the conventional view, in which
the accretion rate matches the return rate's $t^{-5/3}$ time-dependence, this much mass would have
been accreted by $\tau \simeq 1.6$, and 80\% would have been accreted by $\tau \simeq 12$.

\subsection{Cooling time}

One of the key assumptions of our calculation was that the flow is adiabatic.   We can test the validity of
that assumption by estimating the radiative cooling time $t_{\rm cool} \sim \tau_T H/c$, where $\tau_T$
is the (half-)optical depth to Thomson scattering in the vertical direction.    The cooling time declines
outward, but rather slowly, because $H/r$ varies little with radius.   As a result, the outward
decline in surface density is partially offset by the outward increase in $r/c$.    For our parameters, 
$t_{\rm cool}$ is $\sim 50$--150~yr, or $\sim 5 \times 10^4 t_0$, for $r \lesssim 1000M$.

\section{Discussion}\label{sec:interp}

\subsection{How does the accretion rate onto the black hole vary in time?}

As we have shown, shock deflection removes angular momentum and orbital energy from parts of the stellar debris,
transferring this angular momentum and energy to other parts.   By this process, roughly a third of the bound
mass is forced inward relatively quickly, albeit on a timescale rather slower than the nominal mass-return timescale.
Nearly all the mass that finds its way to $r < 2 R_p$ has, by the end of the simulation, gone all the way through the
inner boundary at $30R_g = 0.3 R_p$.    However, half the inflow through the inner boundary occurs after $\tau \simeq 6$.
Once within $30R_g$, if MHD turbulence develops in the usual way, the inflow time is relatively swift:
\begin{equation}
t_{\rm inflow} \sim 7 \times 10^3 (\alpha/0.1)^{-1} [(H/r)/0.5]^{-2} (r/30R_g)^{3/2} R_g/c
\end{equation}
or $\tau_{\rm inflow} \sim 0.1 (\alpha/0.1)^{-1} [(H/r)/0.5]^{-2}(r/30R_g)^{3/2}$.
Here $\alpha$ is the usual ratio of vertically-integrated stress to vertically-integrated pressure.
Because $t_{\rm inflow}$ of the matter pushed early through the inner boundary is much shorter than the
time it takes to reach that boundary, accretion of this material onto the black hole and arrival at the inner
boundary are nearly instantaneous.

The majority of the bound mass, meanwhile, settles into the flow at much larger distance.   At the characteristic
radius of this portion of the stellar debris, the inflow time is much longer: 
\begin{equation}
t_{\rm inflow} \sim 1.3 \times 10^6 (\alpha/0.1)^{-1} [(H/r)/0.5]^{-2} (r/1000R_g)^{3/2} R_g/c,
\end{equation}
or $\tau_{\rm inflow} \sim 19 (\alpha/0.1)^{-1} [(H/r)/0.5]^{-2} (r/1000R_g)^{3/2}$.
Thus, the majority of the mass arrives at the black hole only after a considerable delay.

We can roughly predict the accretion rate at later times by constructing
a crude approximate description for the action of turbulent MHD stresses.    Suppose that the time
at which matter accumulated at radius $r$ reaches the black hole's ISCO is
\begin{equation}
t_{\rm acc} = t_{\rm arrive} + \left[ \alpha^{-2} (H/r)^{-2} + n_{\rm sat}\right] (r/R_g)^{3/2} R_g/c .
\end{equation}
The rate at which matter accretes onto the black hole is then roughly
\begin{equation}
{dM \over dt} = 2\pi r \Sigma[r(t_{\rm acc})] {dr \over dt_{\rm acc}},
\end{equation}
where we have identified the time of observation $t$ with $t_{\rm acc}$.

Several parameters govern the behavior of this model.   Our hydrodynamic simulation results suggest that the time at
which matter settles into the flow, $t_{\rm arrive}$, is $\simeq 12t_0$.   Although, as shown by Figure~\ref{fig:mr_net},
$M(<r)$ approaches its final value at most radii by $\tau \simeq 5$, the data displayed in Figure~\ref{fig:sigma_rho_evol} demonstrate
that the motion of the matter remains dominated by shocks and non-circular flow until nearly the end of the simulation.  
Because this is a newly-formed flow, a finite time is required for the MHD turbulence to reach saturation; we call
this $n_{\rm sat}$ dynamical times.  Numerous simulations, both of shearing boxes and global disks, have shown that
$\sim 10$~orbital periods are generally required for the turbulence to reach saturation; on this basis, we estimate
$n_{\rm sat} \simeq 50$.   In a steady-state disk, the mean inflow time is $\alpha^{-2} (H/r)^{-2}(r/R_g)^{3/2} R_g/c$.
In this context,  the saturation level of the MHD turbulence, and therefore the best value of $\alpha$ for estimates, is more
uncertain than usual.   Even at $\tau \gtrsim10$, the accretion flow is still not entirely circular or axisymmetric.   More
importantly, it is intrinsically time-dependent.    We therefore suppose only that $\alpha \sim 0.01$--0.1.

\begin{figure}
\begin{center}
\includegraphics[width=140mm]{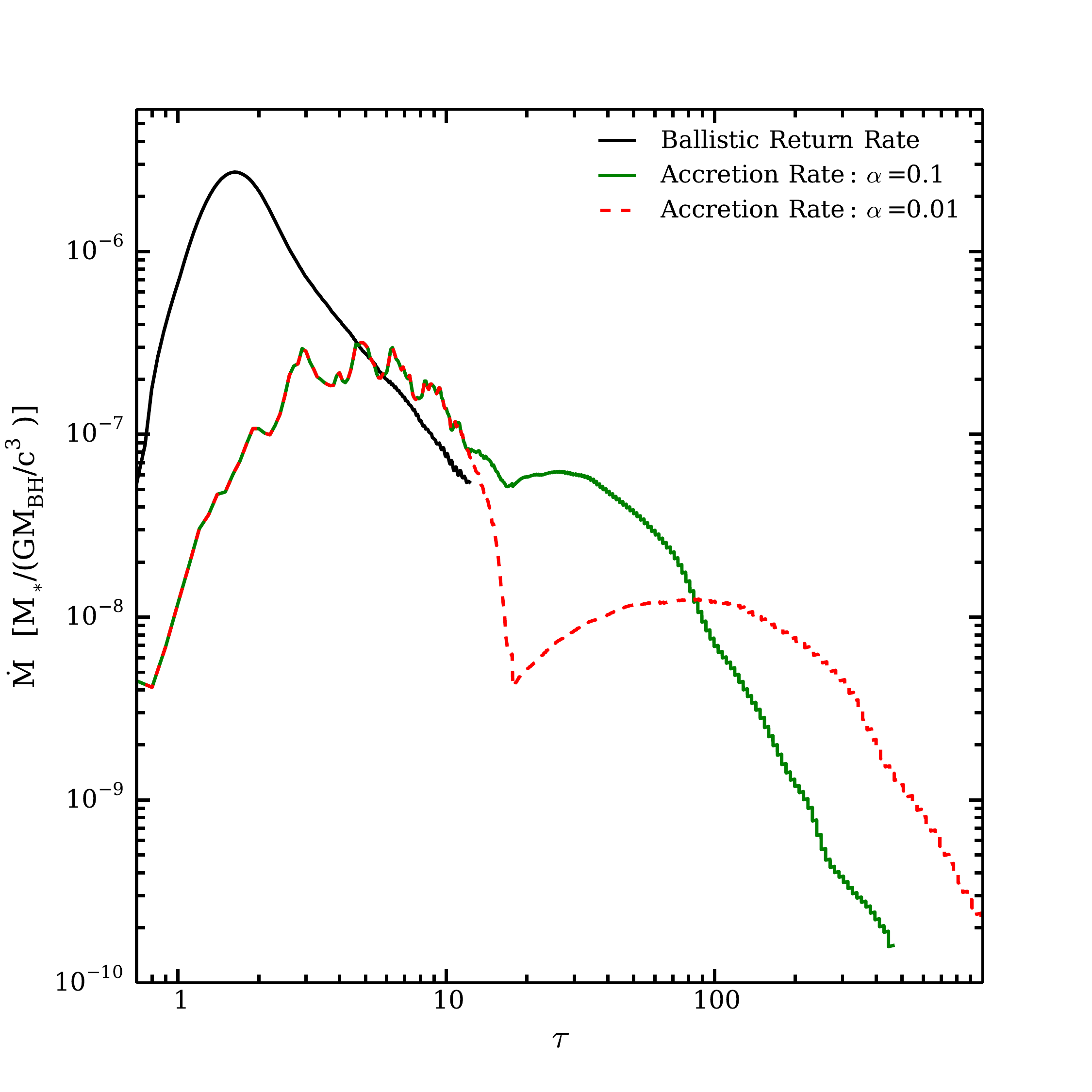}
\end{center}
\caption
{
Accretion rate onto the black hole.   The black curve shows the usual assumption, in which the accretion
rate is the same as the ballistic return rate.  The red and green curves show the sum of the rate at which matter
passes through the inner boundary in our simulation and the accretion rate predicted by our approximate model.
They differ in their choices of $\alpha$, 0.1 for the green curve, 0.01 for the red curve.
}
\label{fig:lightcurve}
\end{figure}

The predicted accretion rate derived from such a model is shown in Figure~\ref{fig:lightcurve}.     The peak accretion rate is due to the
early forcing of matter inward.   It is reached rather later ($\tau \simeq 3$ rather than $\tau \simeq 1$)
than has generally been assumed and is maintained rather longer (until $\tau \simeq 8$).    The value of the
peak accretion rate is $\simeq 0.1\times$ the peak in the ballistic mass-return rate, the usual estimator of the accretion
rate.    This order of magnitude reduction in the maximum mass-return rate is due in part to the fact that the width of the
peak is greater by a factor $\simeq 5$ and in part to the fact that the mass arriving during the peak is only about 1/2 the
mass delivered at the peak ballistically.   After $\tau \simeq 8$,
accretion of the early-arriving material drops off very quickly; by $\tau \simeq 10$--15 (the exact value depending
on $\alpha (H/r)^2$), accretion of the material initially placed at larger distance dominates.   The accretion rate
at late times also depends on $\alpha (H/r)^2$, but is likely to be at least a factor of several below the peak accretion
rate resulting from the early accretion.    On the other hand, once the later mechanism becomes established,
the accretion rate can remain roughly constant for a significant time, $ \sim 30$--100 in $\tau$ units (multiples
of $t_0$, the orbital period of the most tightly-bound tidal stream).
At truly long times, the accretion rate resumes a power-law drop-off whose slope depends on the radial
surface density profile at the time the mass settles into the flow and on the stress parameters.

If the accretion flow is radiatively efficient, the lightcurve should closely mirror the accretion rate.   It is convenient
in this context to translate the dimensionless units of Figure~\ref{fig:lightcurve} into the dimensionless units of
ratio to the Eddington accretion rate $\dot M_E$: the conversion factor is such that $10^{-7} M_*/(GM_{\rm BH}/c^3)
 = 5.57 \times 10^7 (\eta/0.1)\dot M_E$, where $\eta$ is the radiative efficiency of accretion in rest-mass units.
Thus, for the parameters of our simulation, the peak accretion rate is extremely super-Eddington,
$\sim 1 \times 10^8 \dot M_E$.   For such a very large accretion rate, photon trapping effects may severely
limit the flow's radiative efficiency \citep{Begelman1979, Abramowicz1988}.   On the other hand, magnetic buoyancy effects
\citep{Blaes2011} may be able to carry heat to the outer layers rapidly enough to restore some or perhaps all of
the system's radiative efficiency \citep{Jiang2014}.

Despite these uncertainties, it is worthwhile considering what our approximate model for the accretion rate predicts in
the event of efficient radiation.   Particularly if accretion stresses are relatively large (e.g., the green curve in Fig.~\ref{fig:lightcurve}),
the form of the lightcurve remains one in which there is a rapid rise to a peak followed by a long, roughly power-law
decay.    The shape of this decay is, however, not exactly the conventional $t^{-5/3}$ decline.   Although there
are periods when it is reasonably well described by a power-law with approximately this index, there is also
a period during which the accretion rate is roughly constant at a level that is a fraction of
the peak value.   The internal accretion rate, parameterized by $\alpha (H/r)^2$, plays the greatest role in
determining the level at which this flat section occurs and its duration.    To the extent that photon-trapping plays
a role, the peak of the lightcurve would be depressed and the early phase of the lightcurve flattened.

\subsection{Extrapolation to main sequence star disruptions and larger $M_{\rm BH}$}

The parameter regime of our simulation---a white dwarf passing close to a $500 M_{\odot}$ black hole---refers
to a minority subclass of possible tidal disruption events.   It is worthwhile
examining the extent to which its implications can be extrapolated to other, more common, parameters.    The
most natural way to make the connection is through $R_p$ in gravitational units.   We studied an event in
which a white dwarf has $R_p = R_t = 100R_g$; if a main sequence star of mass $M_{\rm MS}$
were to have an encounter in which this were also true, the black hole mass would be
\begin{equation}
M_{\rm BH}^\prime = M_{\rm BH} (M_{\rm WD}/M_{\rm MS})^{1/2} (R_{\rm MS}/R_{\rm WD})^{3/2}.
\end{equation}
Because main sequence star radii are $\sim 100 \times$ larger than white dwarf radii, the black hole for
the corresponding main sequence tidal disruption would be $\sim 1000\times$ larger than the black hole
in our simulation; to be specific, $M_{\rm BH}^\prime = 3 \times 10^5 M_{\rm MS}$ because
$R_* \propto M_*$ for stars with mass $\sim M_{\odot}$ on the main sequence.

Although we have matched $R_p/R_g$, the orbital scale for the most tightly-bound tidal streams will
not be the same: it is $a_{\rm min}/R_g \sim (M_{\rm BH}^\prime/M_*)^{1/3} (R_t/R_g)$.   Thus, in gravitational units,
$a_{\rm min}$ for a main sequence disruption is an order of magnitude larger than for a white dwarf
disruption.   At early times, the speeds of shocks~2 $\sim (a_{\rm min}/R_g)^{-1/2}$, so the heat generated per
unit mass passing though the shocks would fall by about that same single order of magnitude, although as they
extend inward at later times, this effect will be somewhat mitigated.

On the other hand, for fixed $R_p/R_g$, the photon diffusion time in shocked material  scales
$\propto a_{\rm min}^{-1}$ in physical units,
or $\propto {M_{\rm BH}^{\prime}}^{-4/3}$.    We would then expect a decrease in the cooling time by a factor
$\sim 10^4$, to $\sim 3 \times 10^5$~s.     The relevant standard of comparison for the cooling time is $t_0$, and it scales
$\propto {M_{\rm BH}^{\prime}}^{3/2}$ for fixed $R_p/R_g$.    Consequently, the ratio $t_{\rm cool}/t_0$ is extremely
sensitive to black hole mass, $\propto {M_{\rm BH}^{\prime}}^{-17/6}$.    For
this reason, the adiabatic approximation, although well-supported for our parameters, breaks
down for $M_{\rm BH}^\prime \gtrsim 2 \times 10^4 M_{\odot}$.   Post-shock cooling when the black hole
mass is larger should then be significant, possibly further postponing the erasure of density inhomogeneities in the
flow.   Quick post-shock cooling could also create a precursor signal preceding the main
flare that takes place when matter begins to accrete onto the black hole.    However, the available energy
per unit mass is at most $\sim c^2 R_g/a_{\rm min}$, and this will always be small compared to the black hole
accretion energy release unless nearly all of the photons created near the black hole are trapped in the inflow.

Our most important result is impervious to these changes of scale.    The orbital period of the most tightly-bound
matter $t_0$ determines {\it all} aspects of the inflow: the ballistic dynamics of the tidal streams, the orbital period at the
characteristic radius of shocks~2 and 3, and the inflow time from that characteristic radius ($t_{\rm inflow}
\sim \alpha^{-1} (H/r)^{-2} t_0$).    Because the mass of the disrupted star is $\sim M_{\odot}$ whether it is a white
dwarf or on the main sequence, the time-dependence of the accretion rate depends only on this single parameter.
Therefore, by plotting the accretion rate in terms of time in units of $t_0$, the time-dependence shown in
Figure~\ref{fig:lightcurve} should apply reasonably well to {\it all TDEs}, with only minor adjustments due to
variations in $R_p/R_t$, the internal structure of the disrupted star, etc.

Thus, all that is necessary to predict the time-dependence of accretion rate during a main sequence TDE
is to reinterpret $\tau$ in terms of the appropriate $t_0$.    Comparing our results to the predictions of the
classical theory, we find that we, too, expect a sharp rise followed by an extended, roughly power-law
decay.   However, there are two contrasts.    For a given set of parameters ($R_p/R_t$, $M_{\rm BH}$, $M_*$),
the peak accretion rate is $\simeq 10\times$ smaller, and the relation between the duration of the peak
is about $5\times$ longer.    The factor of ten diminution in maximum accretion rate means that it becomes
 $\simeq 80 M_*^{(1+3\xi)/2} M_{BH,6}^{-3/2}$ in Eddington units.    This reduction in the accretion rate
relative to Eddington should also lead to a weakening of photon-trapping.    Because $t_0 \propto (M_{\rm BH}M_*^{1-3\xi)})^{1/2}$,
the geometric mean of the black hole and stellar masses inferred from a measured timescale is $\simeq 5\times$ smaller
than the one predicted by the classical theory.

\section{Summary}

When a star is tidally disrupted by a black hole, its matter is dispersed into thin streams that travel far from the
place where the star was torn apart before returning to the vicinity of the black hole.    Before those streams
can accrete onto the black hole and generate the energy for a flare, their orbital energy must be dissipated,
much of the heat generated must be lost, and their angular momentum must be reduced to less than that
supporting an orbit at the black hole's ISCO.
In this paper, we have focused on the first step---how the orbital energy is dissipated---and we have done so
by consciously ignoring MHD effects, which almost certainly are important to the last step (angular momentum
transport) and possibly even to the middle step (heat loss: see \cite{Blaes2011,Jiang2014}).

Previous work has emphasized the role of what we have named ``shock~1" (also called the ``nozzle" shock),
a shock formed near the pericenter of the star's orbit, where the various tidal streams converge.    We have
identified a shock system, shocks~2 and 3, that appears near the apocenter of the streams' orbits.
These shocks are created by an intersection between a stream that has just swung through pericenter and
another stream that is falling toward its pericenter.   If the orbits of the two streams have a modest ($\sim 10^\circ$)
contrast in apsidal angle, the angle between their velocities at the point where their orbits intersect
can be much larger, $\sim 90^\circ$.    We have shown
that just such a swing in apsidal angle is created by relativistic effects during the disruption, particularly
the relativistic precession in apsidal angle of the star's trajectory.   Because $R_t \sim O(10) R_g$ is typical,
these modest contrasts in apsidal angle can be expected generically in stellar tidal disruptions.

All three shocks dissipate orbital energy, although the amount they dissipate is small in rest-mass terms,
$\sim 5 \times 10^{50}$~erg altogether in this simulation, $\sim 10^{-3}\times$ the rest-mass energy of
the stellar mass bound to the black hole.     Although shocks~2ab contribute the largest share of the heating,
shock~1's share is smaller by only a factor of a few, and part of this shock pair moves in toward the location of
shock~1 during the time that their heating rate is greatest.    We therefore expect the amount of heating in
these shocks to scale with $R_g/R_p$, a number that is relatively insensitive to black hole mass.

In addition,  by deflecting the tidal streams, the shocks substantially widen what was
initially a rather narrow specific angular momentum distribution, causing some material to plunge more directly
toward the black hole while other material follows a more circular orbit at larger distance.   This mechanism can
cause an interesting fraction of the mass to reach radii a factor of a few smaller than $R_p$ without any of
the usual MHD internal stresses.  The broader angular momentum distribution also leads to a wider range of
orbital paths for gas passing through the pericenter region for a second or third time, significantly broadening
the front associated with shock~1, extending it both inward and outward radially.

Previous work had also assumed that the returning tidal streams would quickly join an accretion disk whose
outer edge would lie at the radius, $2R_p$, where the specific angular momentum to support a circular orbit matched
the specific angular momentum of the star's trajectory.   However, we find that shocks~2 and 3 lead to such a
build-up of material in the vicinity of $r \sim a_{\rm min}$, the semi-major axis of the most tightly-bound material,
that the outer edge of the accumulated material is on that scale, rather than the considerably smaller $R_p$.
Moreover, there are significant structural contrasts between this ``disk" and a classical accretion disk.   Even
long after the great majority of the star's bound mass has returned to pericenter at least once, the structure
is strongly asymmetric, and the fluid orbits are rather eccentric.

These results have strong implications for the interpretation of TDE observations.
Arrival of matter at radii close enough to the black hole to release a substantial amount of energy appears to
be subject to several kinds of delays.    A significant minority of the star's bound mass is pushed well within
the star's pericenter, but only after a time $\sim 5t_0$, where $t_0$ is the orbital period of the most tightly-bound matter,
whereas classical expectations about tidal disruptions assumed that much of the star's mass would reach the
black hole only 1--$2t_0$ after the star passed through pericenter.    Accretion of the majority of the star's
bound mass is delayed by a considerably longer time because, at its scale $\sim a_{\rm min}$, inflow driven
by the usual internal stresses associated with MHD turbulence takes a much longer time.   Although the classical
scaling of characteristic timescale $\propto (M_{\rm BH} M_*)^{1/2}$ remains appropriate, these several sources of
delay insert a proportionality constant in this relation
$\sim 5 \times$ larger than generally estimated.   As a result, the product of $M_{\rm BH}$ and $M_*$ inferred
from a particular lightcurve would be an overestimate by $\sim 25$, and the peak accretion rate in Eddington
units is reduced by a factor $\sim 10$.  However, this suggestion requires confirmation (or refutation) by a future MHD simulation.
It is conceivable that MHD effects (e.g., stresses associated with magnetic field amplification by fluid shear)
could also influence the flow even before formation of anything resembling an accretion disk.

\acknowledgements{This work was partially supported by NSF grant AST-1028111 and NASA/ATP grant
NNX14AB43G (JHK), and NSF grants OCI-0725070, OCI-0832606, AST-1028087, and PHY-1125915 (SCN).
Additional support was received from an ERC advanced grant on GRBs (TP) and by the I-CORE 
Program of the Planning and Budgeting Committee and The Israel Science Foundation grant No 1829/12 (TP).
This research was also supported in part by the National Science Foundation under Grant No. NSF PHY11-25915
to the Kavli Institute for Theoretical Physics, where much of this work was initiated.}

\bibliography{ms}

\end{document}